\documentclass[aps,pra,twocolumn,showpacs]{revtex4}
\usepackage{color,graphicx,hyperref,ulem}
\usepackage{amsmath, amssymb}

\begin{document}

\title{Homodyne detection with on-off detector systems}

\author{T. Lipfert}\affiliation{Arbeitsgruppe Theoretische Quantenoptik, Institut f\"ur Physik, Universit\"at Rostock, D-18051 Rostock, Germany}
\author{J. Sperling}\email{jan.sperling@uni-rostock.de}\affiliation{Arbeitsgruppe Theoretische Quantenoptik, Institut f\"ur Physik, Universit\"at Rostock, D-18051 Rostock, Germany}
\author{W. Vogel}\affiliation{Arbeitsgruppe Theoretische Quantenoptik, Institut f\"ur Physik, Universit\"at Rostock, D-18051 Rostock, Germany}

\pacs{42.50.Ar, 03.65.Wj, 85.60.Gz}
\date{\today}

\begin{abstract}
	Phase-sensitive properties of light play a crucial role in a variety of quantum optical phenomena, which have been mostly discussed in the framework of photoelectric detection theory.
	However, modern detection schemes, such as arrays of on-off detectors, are not based on photoelectric counting.
	We demonstrate that the theory of homodyning with such click counting detectors can be established by using a proper detection model.
	For practical applications, a variety of typically occurring imperfections are rigorously analyzed and directly observable nonclassicality criteria are studied.
	Fundamental examples demonstrate the general functionality of our technique.
	Thus, our approach of homodyne detection with on-off detector systems is able to bridge the gap between imperfect detection and the phase resolution demands for modern applications of quantum light.
\end{abstract}

\maketitle

\section{Introduction}\label{sec:1}
	Phase-sensitive measurements have a fundamental impact on uncovering quantum properties of radiation fields~\cite{MW95,VW06,A13}.
	Applications of quantum effects can be found in the vast fields of quantum information science and quantum metrology~\cite{KWM00,PDFEPW10,PADLA10,CIDE14}.
	Those are typically studied in terms of homodyne measurements with high intensities and the photoelectric detection theory~\cite{WVO99,LR09}.
	However, applications of light quanta in the single- or few-photon domain faces some flaws as realistic photon-number resolving detectors are often not available~\cite{SV11}.

	The nonclassicality of light is usually determined with reference to classical coherent states~\cite{TG65,M86}.
	More precisely, any quantum state can be represented in the basis of these classical states by the Glauber-Sudarshan $P$ phase-space representation~\cite{G63,S63}.
	Whenever the $P$ function is nonnegative, it can be interpreted as a probability distribution of a classical ensemble of electromagnetic waves.
	If the $P$ function exhibits negativities, no such interpretation can be made, and the state of light is therefore a nonclassical one.
	Hence, the categorization in classical or nonclassical radiation fields plays a crucial role.

	In particular, phase dependent quantum phenomena are often studied using interferometric measurement schemes such as homodyne detection~\cite{NFM91,SBRF93,SBCRF93,FS93}.
	In a homodyning setup, a signal (SI) is superimposed on a beam splitter with a local oscillator (LO), with a controllable phase.
	This is usually done with a strong LO and formulated in terms of the photoelectric detection theory.
	Rigorous analysis for low LO intensities turns out to be more involved but can be analyzed and implemented~\cite{VG93,MBAR95,DBJVDBW14}.
	In addition, homodyne detection serves as a foundation for quantum state reconstruction techniques~\cite{WVO99,LR09,L05} for uncovering all nonclassical features of light.

	Hand in hand with the quantum nature of the photon comes its particle nature, which may be confirmed by antibunching effects~\cite{KM76,WC76,KDM77}.
	Thus, to gain further knowledge of the nature of light, the generation and analysis of states in a few-photon regime has gained more and more importance~\cite{BC10,EFMP11}.
	This has brought up the necessity for adequate detectors.
	For example, hybrid detectors~\cite{ALBPMHM13}, superconducting detectors~\cite{SSTT13,MVSHLGVBSMN13}, and click-counting detectors~\cite{RHHPH03,WDSBY04,ZABGGBRP05,JDC07,BGGMPTPOP11,MMDL12}, consisting of multiple avalanche photodiodes (APDs) in the Geiger mode, may meet this demand.
	The latter ones offer a nonlinear, but well-defined, statistics~\cite{FJPF03,BDFL08,SVA12_85}.
	Quantum correlations in the few photon domain can be inferred with such click detectors~\cite{BDFL08,ABA10,ALCS10,DYSTS11,HSRHMSS14,AJBGPHB14}, even without the need for additional data processing~\cite{SBVHBAS15}.
	To get the full quantum picture, it is desirable to investigate both particle- and phase-dependent phenomena simultaneously.
	In Ref.~\cite{SVA15}, a step towards a theory of phase-resolving click counting was made.
	We aim at extending this investigation by giving a more general approach and lay the foundation for further applications, such as quantum state reconstruction~\cite{LSV15}.

	In the current contribution, we formulate the theory for general homodyne detection measurements with click detectors.
	This includes balanced and unbalanced detection with four or more port homodyning setups.
	The verification of nonclassical features is formulated in terms of measured second order correlations, and the straightforward generalization to higher-order correlations is outlined.
	We rigorously investigate the influence of imperfections in such measurement scenarios, which includes, e.g., the impact of detector imperfections as well as LO fluctuations and mode mismatch between LO and SI.
	Moreover, the identification of quantum correlations between multiple SI fields is shown using several homodyne detector settings.

	This paper is structured as follows.
	In Sec.~\ref{sec:2} we give an overview of the treatment of click counting detectors, reviewing results achieved on this topic so far.
	The click detection in general four port homodyning schemes is derived in Sec.~\ref{sec:3}, including the cases of unbalanced and balanced scenarios.
	The influence of imperfections is studied in Sec.~\ref{sec:noise}.
	Multi-port homodyne detection schemes are treated in Sec.~\ref{sec:4} by means of the example of a balanced eight-port setup.
	The discourse on phase-sensitive click counting is then completed in Sec.~\ref{sec:5}, where we discuss homodyne correlation measurements between multiple signal fields.
	Finally, we summarize and conclude in Sec.~\ref{sec:6}.

\section{Click-Counting Theory}\label{sec:2}

\subsection{Click counting versus photoelectric counting}
	Most commonly, when detection processes of light are discussed in quantum optics, the theory of photoelectric measurement is examined.
	For the single mode case, the resulting photoelectric counting statistics reads~\cite{MW95,VW06,A13}
	\begin{equation}\label{eq:PhotoelectricStatistics}
	 	p_n=\langle{:} \frac{(\eta \hat  n +\nu)^n}{n!} {e}^{-(\eta \hat n  +\nu)}{:}\rangle ,
	\end{equation}
	where  ${:}\cdots{:}$ is the normal ordering prescription, $\eta$ is the quantum efficiency, and $\nu$ is the dark count rate~\cite{STG08}.
	The photoelectric statistics is (i) the true photon number statistics for a perfect detection scenario, $\eta=1$ and $\nu=0$, and (ii) a true Poisson statistics for coherent light.

	\begin{figure}[ht]
	 	\vspace{0.6233cm}\includegraphics[width=5.64cm]{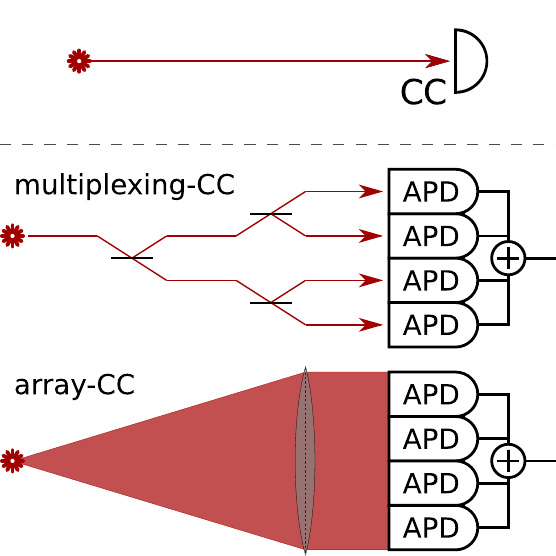}
	 	\caption{
			(Color online)
	 		The top illustration shows the integrated symbol ``CC'', which we will further use for click counting detectors of any type.
	 		Two realizations of click detectors with $N=4$ APDs are additionally shown.
	 		In the spatial multiplexing setup (middle scheme) the incident light is equally divided by multiple $50{:}50$ beam splitters.
	 		Time-bin multiplexing is an equivalent realization.
	 		In the detector array scenario (bottom scheme), an array of APDs is equally illuminated.
	 	}\label{Click_Fig_Types}
	\end{figure}

	Alternatively, detector systems based on APDs can be applied.
	In such a configuration, each APD acts as an on-off detector~\cite{BC10,EFMP11}.
	Realizations of detector systems consisting of multiple APDs are detector arrays or multiplexing setups (cf. Fig.~\ref{Click_Fig_Types}).
	Time-bin multiplexing is related to the spatial multiplexing setup (cf., e.g.,~\cite{RHHPH03,FJPF03,ALCS10,ASSBW03}).
	Experimental characterizations of such click detector systems have been performed in Refs.~\cite{LFCPSREPW09,FLCEP09}.

	A fundamental approach to retrieve a kind of photon number resolution is to equally distribute the incident light onto $N$ APDs.
	This yields a total number of $k$ clicks, $0\leq k\leq N$.
	The measured statistics of such click detector devices can be described with a quantum version of a binomial statistics~\cite{SVA12_85}:
	\begin{align}\label{eq:ClickStatistics}
		c_k=\langle {:}
			\binom{N}{k} \left(e^{-(\eta \frac{\hat n}{N}+\nu)}\right)^{N-k} \left(\hat 1-e^{-(\eta \frac{\hat n}{N}+\nu)}\right)^{k}
		{:}\rangle\text{,}
	\end{align}
	where $N$ is the number of APDs and $c_k$ gives the probability to measure $k$ clicks.
	The mean value of this statistics may be given by the expectation value of the operator $\hat\pi$:
	\begin{align}
		\langle\hat \pi\rangle=\sum\limits^{N}_{k=0}kc_k,
		\text{ with } \hat\pi=N\left(\hat 1-{:}e^{-(\frac{\eta}{N}\hat n +\nu)}{:}\right).
	 	\label{pi_plain}
	\end{align}
	This operator is a nonlinear function of the photon number operator.
	Note that we use a rescaled version (factor $N$) of the operator introduced in Ref.~\cite{SVA13}.

	Due to the finite nature of the click-counting statistics, $0\leq k\leq N$, all possible events are accessible in experiments.
	In contrast, the photoelectric detection model~\eqref{eq:PhotoelectricStatistics} (only in theory) allows one to have arbitrarily high $n$ values.
	For instance, the prediction of a coherent state $|\alpha\rangle$ yields $p_n\neq0$ for all $n$ values, which cannot be achieved with a finite number of measured events.

	\begin{figure}[ht]
		\centering
		\includegraphics[width=8.7cm]{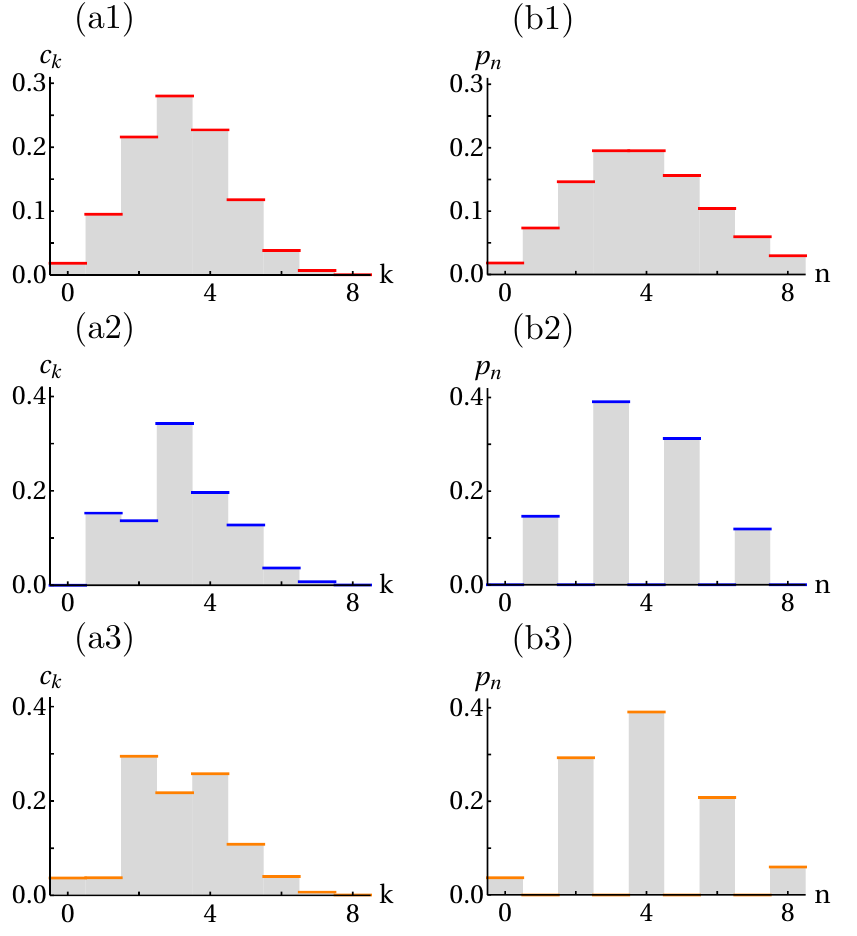}
	 	\caption{
	 		(Color online)
	 		(a1)-(a3) The click statistics as given in Eq.~\eqref{eq:ClickStatistics} for $N=8$ is compared to (b1)-(b3) the corresponding photoelectric counting statistics in Eq.~\eqref{eq:PhotoelectricStatistics}.
	 		The considered states are (top) the coherent state $|\alpha\rangle$, as well as (middle) odd and (bottom) even coherent states~\eqref{eq:EvenOddState}.
	 		For all plots we take $\alpha=2$, $\eta=1$, and $\nu=0$.
	 	}\label{Click_Fig_odd_even_example_const_phase}
	\end{figure}

	Let us give another example to underline the difference between click counting and photoelectric counting theory.
	Three prominent states will serve as our test states throughout this paper.
	One is a classical coherent state $|\alpha\rangle$, and, to illustrate the basic differences from nonclassical light fields, we analyze the even (${+}$) and odd (${-}$) coherent states~\cite{DMM74},
	\begin{align}\label{eq:EvenOddState}
	 	|\alpha_\pm\rangle=\frac{|+\alpha\rangle\pm|-\alpha\rangle}{\sqrt{2[1\pm\exp(-2|\alpha|^2)]}}.
	\end{align}
	In Fig.~\ref{Click_Fig_odd_even_example_const_phase}, we compare the click statistics [(a1)-(a3)] and photoelectric counting statistics [(b1)-(b3)] of these states for a perfect detection scenario, $\eta=1$ and $\nu=0$.
	Let us stress that the click statistics is limited to the displayed plot range, $0\leq k\leq 8=N$, whereas the photoelectric statistics is not.
	Figures \ref{Click_Fig_odd_even_example_const_phase}(a1) and \ref{Click_Fig_odd_even_example_const_phase}(b1), \ref{Click_Fig_odd_even_example_const_phase}(a2) and \ref{Click_Fig_odd_even_example_const_phase}(b2), and \ref{Click_Fig_odd_even_example_const_phase}(a3) and \ref{Click_Fig_odd_even_example_const_phase}(b3) show the coherent, even coherent, and odd coherent states, respectively.
	It is clearly visible that the true photon statistics of the (even)odd coherent state includes only (even)odd photon numbers.
	This feature is not distinct in the corresponding click statistics, even for the studied ideal detection.

\subsection{Moment based nonclassicality probes}
	Based on the variance of the click statistics, it has been demonstrated that it is possible to infer nonclassical features of quantum light, yielding the notion of sub-binomial light~\cite{SVA12_109,BDJDBW13}.
	In general, it was also shown that higher-order ordered moment criteria can be used to asses these criteria~\cite{SVA13}.
	For this reason the matrix of $K$th-order moments of the click statistics was defined as
	\begin{align}\label{eq:MoM}
		\boldsymbol M^{(K)}=&\left(\langle{:}\hat\pi^{m+m'}{:}\rangle\right)_{m,m'=0}^{K/2},
	\end{align}
	where the even integer $K$ satisfies $K\leq N$.
	If we do not restrict ourselves  a particular moment order, we skip the upper index and write $\boldsymbol M$.
	For classical radiation fields this matrix is non-negative, $\boldsymbol M\geq 0$, and a violation of this property uncovers quantum light.

	It might be more convenient to formulate the modified matrix of moments in terms of
	\begin{align}
		\langle{:}(N-\hat\pi)^{m+m'}{:}\rangle=\langle{:}\left(Ne^{-(\eta\hat n/N+\nu)}\right)^{m+m'}{:}\rangle 
	\end{align}
	or central moments
	\begin{align}
		\langle{:}(\Delta\hat\pi)^{m+m'}{:}\rangle=\langle{:}(\hat\pi-\langle{:}\hat\pi{:}\rangle)^{m+m'}{:}\rangle.
	\end{align}
	The resulting ways for determining nonclassicality are identical for all the different formulations of matrices of click counting moments, which is proven in Appendix~\ref{App:MOMexpansion}.
	The non-negativity of all these matrices of moments can be inferred from their principal minors.
	For example, a second-order moment based constraint for classical light is
	\begin{align}
		0\leq &\langle{:}[\Delta\hat\pi]^2{:}\rangle=\det\boldsymbol M^{(2)}
		\\\nonumber&=\det\begin{pmatrix}
			1 & \langle{:}\hat\pi{:}\rangle \\ \langle{:}\hat\pi{:}\rangle & \langle{:}\hat\pi^2{:}\rangle
		\end{pmatrix}
		=\det\begin{pmatrix}
			1 & 0 \\ 0 & \langle{:}(\Delta\hat\pi)^2{:}\rangle
		\end{pmatrix}
		\\\nonumber&=\det\begin{pmatrix}
			1 & \langle{:}(N-\hat\pi){:}\rangle \\ \langle{:}(N-\hat\pi){:}\rangle & \langle{:}(N-\hat\pi)^2{:}\rangle
		\end{pmatrix}.
	\end{align}
	If, in addition, we consider two click-counting detectors, we can also formulate cross-correlation criteria by the corresponding minors~\cite{SVA13}.
	For example, the second order criterion is
	\begin{align}
		0\leq \langle{:}(\Delta\hat\pi_1)^2{:}\rangle\langle{:}(\Delta\hat\pi_2)^2{:}\rangle-\langle{:}\Delta\hat\pi_1\Delta\hat\pi_2{:}\rangle^2,
	\end{align}
	where each click detector ($i=1,2$) is characterized by the operator $\hat \pi_i$ [cf. Eq.~\eqref{pi_plain}].
	These cross correlations have been recently used to experimentally uncover quantum correlations between two light beams~\cite{SBVHBAS15}.

	Finally, it is worth mentioning that the moments for any number of click-counting devices can be directly obtained from the measured joint click statistics $c_{k_1,k_2,\ldots}$~\cite{SVA13}.
	This yields joint moments of the form
	\begin{align}\label{eq:samplingformula}
	 	\langle{:}\prod_i\hat\pi_i^{m_i}{:}\rangle=\prod_i N_i^{m_i}\sum_{k_1=m_1,k_2=m_2,\ldots}^{N_1,N_2,\ldots} \prod_i\frac{\binom{k_i}{m_i}}{\binom{N_i}{m_i}}c_{k_1,k_2,\ldots},
	\end{align}
	which makes the matrix of click-counting moments approach easily accessible.
	Based on these moments, a construction of higher-order or cross-correlation nonclassicality conditions can be formulated for any number of click detector systems; see also~\cite{SVA13}.

\subsection{Outline}
	In summary, the click counting approach is a directly applicable technique to infer phase-insensitive quantum properties of radiation.
	The click statistics~\eqref{eq:ClickStatistics} itself can be obtained from the photoelectric statistics~\eqref{eq:PhotoelectricStatistics} only via a non bijective mapping from the infinite number of photon numbers to the finite spectrum of click counts (cf. Ref.~\cite{SVA12_85}).
	Due to this feature, it is evident that the implementation of click detectors, in the theory of phase-sensitive measurement, has to be carried out with great care.
	In the same way, nonclassicality probes have to be adjusted to avoid misleading interpretations of the probed state of light.
	In the remainder of this work, we will perform such an analysis for a number of established homodyning schemes in quantum optics.

\section{Four port homodyne detection}\label{sec:3}
\begin{figure}[ht]
	 	\includegraphics[width=4.235cm]{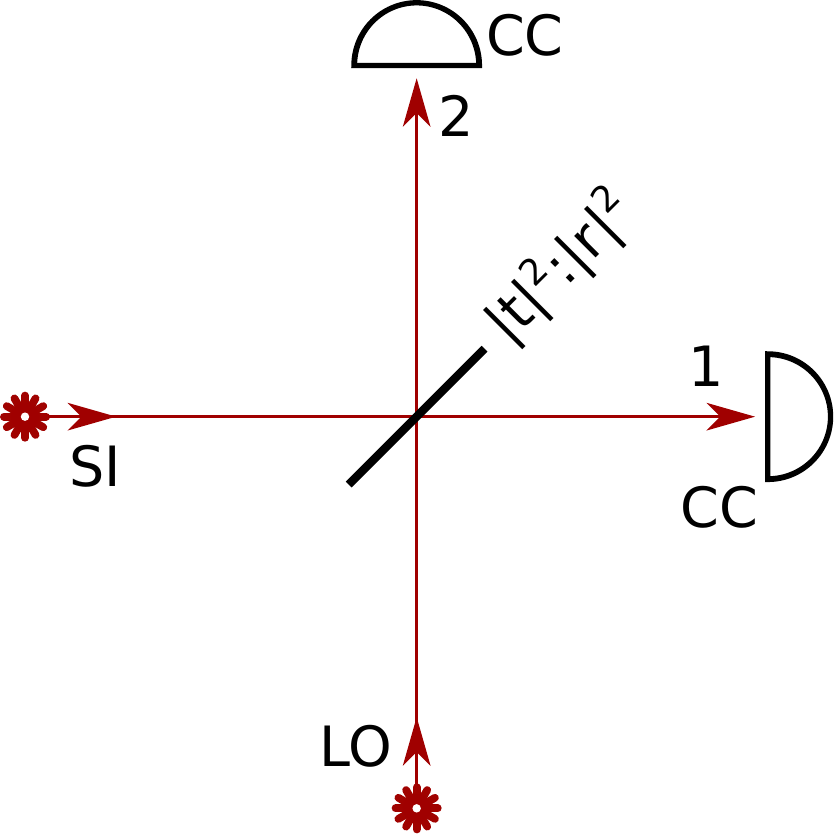}
	 	\caption{
	 		(Color online)
	 		The homodyne four-port scheme.
	 		The LO, having a controllable phase, and the SI are superimposed on a $|t|^2{:}|r|^2$ beam splitter.
	 		The output beams 1 and 2 are each detected with a click detector.
	 	}\label{4PH_scheme}
	\end{figure}

	We will start our treatment of phase sensitive measurement using click detectors with the most prominent setup in homodyne detection: the four port scheme (see Fig.~\ref{4PH_scheme}).
	Applications of this measurement scheme, in terms of photoelectric detection theory, include fundamental applications such as quantum state tomography~\cite{WVO99,LR09,L05}.
	In this measurement scheme, a SI is superimposed with the LO on a beam splitter.
	The latter is described by a transmission coefficient $t$ and a reflection coefficient $r$.
	Let us emphasize that we will describe this setup employing click counting detectors instead of the standard scenario with photoelectric detectors.

	The input-output relation of the beam splitter can be expressed by a unitary transformation matrix,
	\begin{align}
	 	\begin{pmatrix}\hat a_1 \\ \hat a_2\end{pmatrix}
	 	=
	 	\begin{pmatrix}t & r \\ -r^\ast & t^\ast\end{pmatrix}
	 	\begin{pmatrix}\hat a_{\textrm{SI}}\\\hat a_{\textrm{LO}}\end{pmatrix},
	 	\label{4PH_Eq_Unitary}
	\end{align}
	where $|t|^2+|r|^2=1$.
	The click detectors in positions 1 and 2 in Fig.~\ref{4PH_scheme} yield a joint click-counting statistics
	\begin{align}
		c_{k_1,k_2}{=}\langle{:}
			\nonumber
			\binom{N_1}{k_1} \left(e^{{-}\eta_1 \frac{\hat n_1}{N_1}{-}\nu_1}\right)^{N_1-k_1} \left(\hat 1{-}e^{{-}\eta_1 \frac{\hat n_1}{N_1}{-}\nu_1}\right)^{k_1}
			\\\times
			\binom{N_2}{k_2} \left(e^{{-}\eta_2 \frac{\hat n_2}{N_2}{-}\nu_2}\right)^{N_2-k_2} \left(\hat 1{-}e^{{-}\eta_2 \frac{\hat n_2}{N_2}{-}\nu_2}\right)^{k_2}
		{:}\rangle,\label{eq:JointCC}
	\end{align}
	where $\hat n_i=\hat a_i^\dagger\hat a_i$ is the photon number operator, $\eta_i$ is the quantum efficiency, $\nu_i$ is the dark count rate, and $N_i$ is the number of APDs of the click detector at position $i=1,2$.
	The reference beam is a coherent light field, $|\beta\rangle=|\beta\rangle_{\rm LO}$.
	Thus, we can insert the photon number operators in Eq.~\eqref{eq:JointCC} in the form
	\begin{align}
		\hat n_1=&|t|^2\left(\hat a_{\rm SI}+\frac{r}{t}\beta\right)^\dagger\left(\hat a_{\rm SI}+\frac{r}{t}\beta\right),\\
		\hat n_2=&|r|^2\left(\hat a_{\rm SI}-\frac{t^\ast}{r^\ast}\beta\right)^\dagger\left(\hat a_{\rm SI}-\frac{t^\ast}{r^\ast}\beta\right).
	\end{align}
	Note that this form corresponds, up to a scaling, to displaced photon number operators,
	\begin{align}
	 	\hat n(\gamma)=(\hat a_\text{SI} -\gamma)^\dagger(\hat a_\text{SI}-\gamma),
	\end{align}
	where $\hat n=\hat a_{\rm SI}^\dagger\hat a_{\rm SI}$ is the photon number operator of the SI.
	Eventually, the full click-counting statistics of a four port detection scheme is
	\begin{align}
		\nonumber c_{k_1,k_2}
		=\langle{:}&
			\binom{N_1}{k_1}\left(e^{-\eta_1|t|^2\hat n(-r\beta/t)/N_1-\nu_1}\right)^{N_1-k_1}
			\\\nonumber\times&\left(\hat 1-e^{-\eta_1|t|^2\hat n(-r\beta/t)/N_1-\nu_1}\right)^{k_1}
			\\\nonumber\times&\binom{N_2}{k_2}\left(e^{-\eta_2|r|^2\hat n(t^\ast\beta/r^\ast)/N_2-\nu_2}\right)^{N_2-k_2}
			\\\times&\left(\hat 1-e^{-\eta_2|r|^2\hat n(t^\ast\beta/r^\ast)/N_2-\nu_2}\right)^{k_2}
		{:}\rangle.
	\end{align}
	This joint click statistics of the general four port homodyne detector may be used to infer nonclassical light fields via moment criteria.
	For example, the normally ordered click-counting variances or cross correlations certify quantumness if
	\begin{align}
		0>&\langle{:}[\Delta\hat\pi_1]^2{:}\rangle=N_1^2e^{-2\nu_1}\langle{:}[\Delta e^{-\frac{\eta_1|t|^2}{N_1}\hat n(-r\beta/t)}]^2{:}\rangle,\\
		0>&\langle{:}[\Delta\hat\pi_2]^2{:}\rangle=N_2^2e^{-2\nu_2}\langle{:}[\Delta e^{-\frac{\eta_2|r|^2}{N_2}\hat n(t^\ast\beta/r^\ast)}]^2{:}\rangle,\\
		0>&\langle{:}[\Delta\hat\pi_1]^2{:}\rangle\langle{:}[\Delta\hat\pi_2]^2{:}\rangle-\langle{:}\Delta\hat\pi_1\Delta\hat\pi_2{:}\rangle^2.
	\end{align}
	In the following, we will focus on two specific four port homodyning schemes with some relevance in quantum optics.

\subsection{Unbalanced detection}\label{sec:3:a}
	Let us first consider the unbalanced measurement~\cite{WV96} with the click detector only in channel 1 of Fig.~\ref{4PH_scheme}.
	That is, contributions of the detector at position 2 are traced out.
	Alternatively, this corresponds to the case $\nu_2=\eta_2=0$, with a joint click counting statistics~\eqref{eq:JointCC}, which has the property that $c_{k_1,k_2}=0$ for $k_2\neq0$.
	For convenience, we may replace the notations $c_{k,0}=c_k$, $N_1=N$, $\eta_1=\eta$, and $\nu_1=\nu$ in this case.
	Thus, for a LO $|\beta\rangle$, we get
	\begin{align}\label{eq:unbalancedCC}
		c_k=&\langle{:}
			\nonumber \binom{N}{k} \left(e^{-\frac{\eta |t|^2}{N} (\hat a_{\rm SI}+r\beta/t)^\dagger(\hat a_{\rm SI}+r\beta/t){-}\nu}\right)^{N-k}
			\\&\times\left(\hat 1{-}e^{-\frac{\eta |t|^2}{N} (\hat a_{\rm SI}+r\beta/t)^\dagger(\hat a_{\rm SI}+r\beta/t){-}\nu}\right)^{k}
		{:}\rangle\\
		=&\langle{:}
			\nonumber \binom{N}{k} \left(e^{-\frac{\eta_t }{N} \hat n(\gamma){-}\nu}\right)^{N-k}
			\left(\hat 1{-}e^{-\frac{\eta_t }{N} \hat n(\gamma){-}\nu}\right)^{k}
		{:}\rangle,
	\end{align}
	using the beam splitter transformation~\eqref{4PH_Eq_Unitary}, with $\eta_t=|t|^2\eta$ being an overall quantum efficiency and $\gamma=-r\beta/t$.
	Recently, it has been shown in Ref.~\cite{LSV15} that such unbalanced homodyne setups reveal nonclassical features in terms of click counterparts of $s$-parametrized quasiprobabilities.

	If we decompose the coherent displacement in the form $\gamma=|\gamma|e^{i\varphi}$,
	the phase-sensitive mean click counts are given as the expectation value of the operator $\hat\pi(\varphi)$ [cf. Eq.~\eqref{pi_plain}].
	For instance, we get moments of the form
	\begin{align}
	 	\langle{:}[N-\hat\pi(\varphi)]^m{:}\rangle=& \left(Ne^{-\nu}\right)^m
	 	\langle{:}e^{-m\frac{\eta_t }{N} \hat n(\gamma)}{:}\rangle.
	 	\label{4PH_Eq_pi1_Dn}
	\end{align}
	Now, nonclassicality criteria can be constructed. 
	For example, we have nonclassical light for the LO phase $\varphi$ if
	\begin{align}
	 	0>\langle:[\Delta \hat\pi(\varphi)]^2:\rangle=N^2e^{-2\nu}\langle{:}\left[\Delta e^{-\frac{\eta_t }{N} \hat n(\gamma)}\right]^2{:}\rangle.
		\label{4PH_Eq_pi_2nd_order_minor_criterion}
	\end{align}
	Higher-order nonclassicality can be identified with the matrix of phase-sensitive click-counting moments, $\boldsymbol M(\varphi)=(\langle{:}\hat\pi^{m+m'}(\varphi){:}\rangle)_{m,m'}$, which is non-negative for classical light, $\boldsymbol M(\varphi)\geq0$.
	For example, the fourth order nonclassicality condition is
	\begin{align}
		0{>}\det\boldsymbol M^{(4)}(\varphi)
		{=}\det\begin{pmatrix}
			1					& \langle{:}\hat\pi(\varphi){:}\rangle		& \langle{:}\hat\pi^2(\varphi){:}\rangle	\\
			\langle{:}\hat\pi(\varphi){:}\rangle	& \langle{:}\hat\pi^2(\varphi){:}\rangle	& \langle{:}\hat\pi^3(\varphi){:}\rangle	\\
			\langle{:}\hat\pi^2(\varphi){:}\rangle	& \langle{:}\hat\pi^3(\varphi){:}\rangle	& \langle{:}\hat\pi^4(\varphi){:}\rangle
		\end{pmatrix}.
		\label{4PH_Eq_pi_4th_order_minor_criterion}
	\end{align}

\begin{figure}[ht]
		\includegraphics[width=6.3cm]{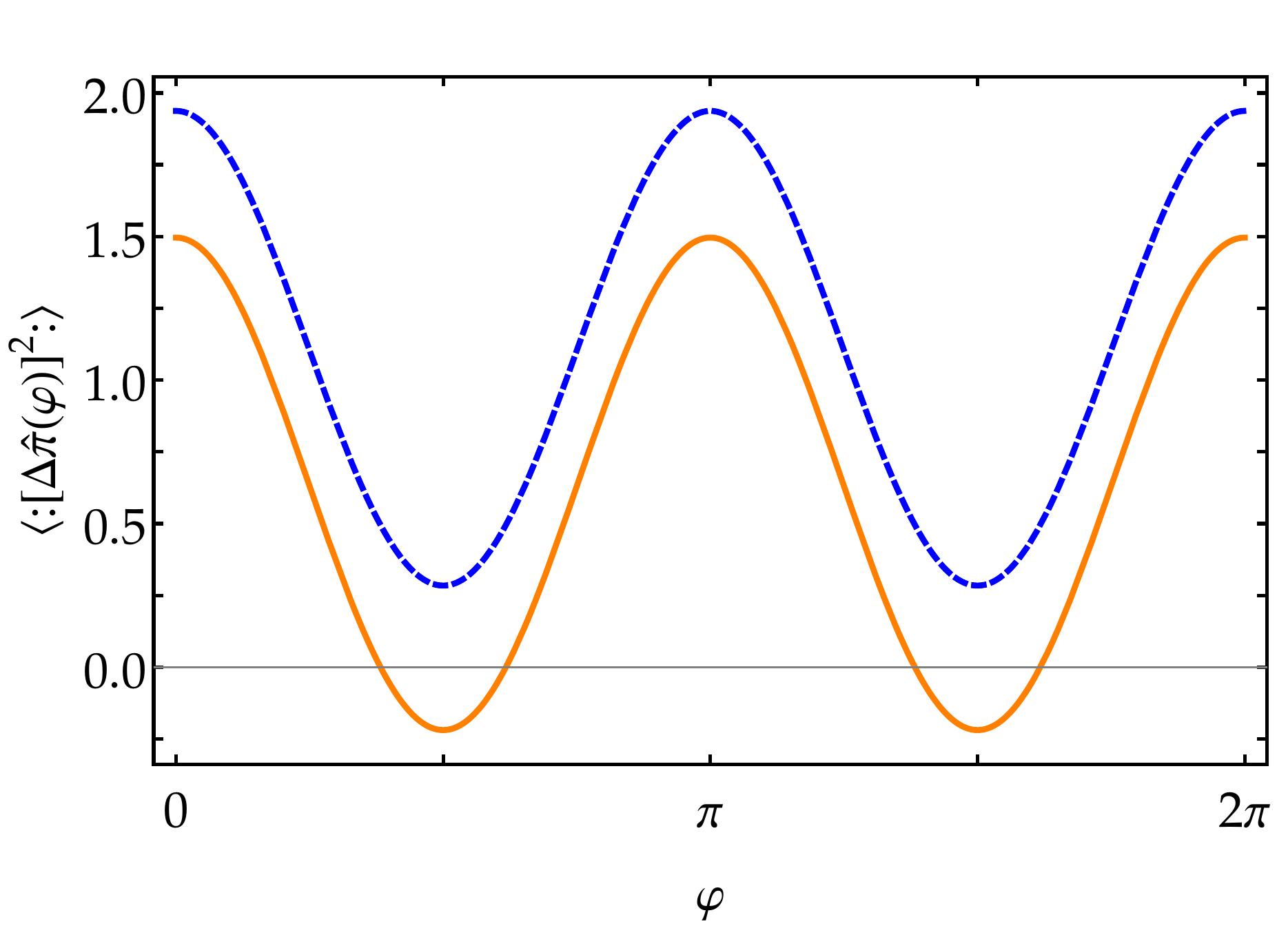}
	 	\caption{
	 		(Color online)
	 		The application of nonclassicality criterion in Eq.~\eqref{4PH_Eq_pi_2nd_order_minor_criterion} is shown as a function of the phase $\varphi$ for $|\alpha_-\rangle$ (dashed line) and $|\alpha_+\rangle$ (solid line), where $\alpha=1$ and $|\beta|=4$.
	 		Negativity of the normally ordered variance determines nonclassicality.
	 		The click detection parameters are $\eta=50\%$, $\nu=0$, $N=8$, and the beam splitter in Eq.~\eqref{4PH_Eq_Unitary} is characterized by $t=4/5$ and $r=3/5$.
	 	}\label{4PH_Fig_pi1_M_V}
	\end{figure}

	Examples of these phase-sensitive variances are given in Fig.~\ref{4PH_Fig_pi1_M_V} for the even and odd coherent states in Eq.~\eqref{eq:EvenOddState}.
	A phase-sensitive verification of nonclassicality can be observed for the even coherent state, while the nonclassicality of the odd coherent state is not revealed by this particular nonclassicality probe.
	The effective quantum efficiency in this unbalanced homodyne detection setup is chosen to be $\eta_t=32\%$.
	For the plot of quantity~\eqref{4PH_Eq_pi_2nd_order_minor_criterion}, the analytic result of the expectation value,
	\begin{align}
		\nonumber &\langle\alpha_\pm|{:}\exp\left[-\lambda(\hat a-\gamma)^\dagger(\hat a-\gamma)\right]{:}|\alpha_\pm\rangle
		\\=&e^{-\lambda|\gamma|^2}\frac{e^{-\lambda|\alpha|^2}\cosh[2\lambda\rm Re(\gamma^\ast\alpha)]}{1\pm e^{-2|\alpha|^2}}
		\\&\pm e^{-\lambda|\gamma|^2}\frac{e^{-(2-\lambda)|\alpha|^2}\cos[2\lambda{\rm Im}(\gamma^\ast\alpha)]}{1\pm e^{-2|\alpha|^2}}\nonumber,
	\end{align}
	was employed.
	It also yields the analytical results for the following discussions.

\begin{figure}[ht]
		\includegraphics[width=8.7cm]{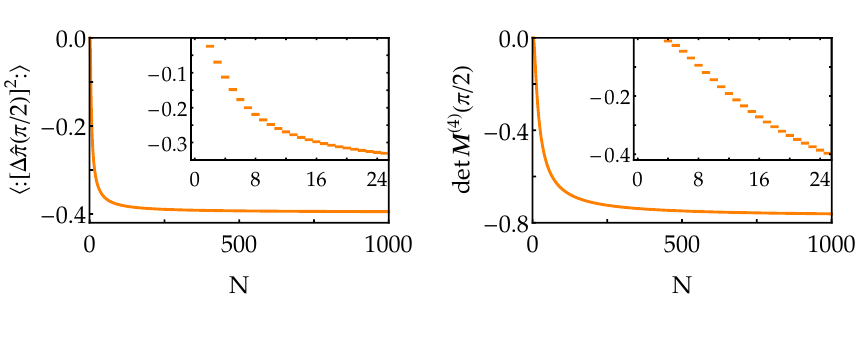}
	 	\caption{
	 		(Color online)
	 		(left) The application of the second-order nonclassicality criterion in Eq.~\eqref{4PH_Eq_pi_2nd_order_minor_criterion} for a fixed phase $\varphi=\pi/2$ and (right) the fourth-order criterion in Eq.~\eqref{4PH_Eq_pi_4th_order_minor_criterion} are shown as a function of the number $N$ of APDs for the even coherent state $|\alpha_+\rangle$ with $\alpha=1$.
	 		The displayed values determine nonclassicality, and the insets illustrate the behavior for small $N$ values.
	 		The click detection parameters are $\eta=50\%$, $\nu=0$, $|\beta|=4$, and the beam splitter in Eq.~\eqref{4PH_Eq_Unitary} is characterized by $t=4/5$ and $r=3/5$.
	 	}\label{4PH_Fig_principalMinors_over_N}
	\end{figure}

	Let us additionally study the verification of nonclassicality for different $N$ values.
	Click detection schemes in Fig.~\ref{Click_Fig_Types} may also employ CCD cameras with a high number of APDs; see, e.g.,~\cite{HPHP05,PHMH12} for theoretical and experimental studies using CCD cameras.
	In Fig.~\ref{4PH_Fig_principalMinors_over_N}, we show the $N$ dependence of the verified nonclassicality for an even coherent state for second- and fourth-order criteria [see Eqs.~\eqref{4PH_Eq_pi_2nd_order_minor_criterion} and~\eqref{4PH_Eq_pi_4th_order_minor_criterion}, respectively].
	The negativities are typically more pronounced for the fourth order criterion (right plot) in comparison with the second order one (left plot).
	It can be observed that higher numbers of APDs are advantageous for those criteria too.
	In the limit $N\to\infty$, the negativities converge to those values which are expected for an unbalanced homodyning scheme with a photoelectric detector.
	
\subsection{Balanced detection}\label{sec:3:b}
	Now we will study the balanced case of the four-port scheme in Fig.~\ref{4PH_scheme}.
	That is, the beam-splitter transformation in Eq.~\eqref{4PH_Eq_Unitary} is specified by $t=r=1/\sqrt 2$.
	In addition, we assume that both detectors have the same characteristics, i.e., $N_1=N_2=N$, $\eta_1=\eta_2=\eta$, and $\nu_1=\nu_2=\nu$.
	On this basis, the joint click-counting statistics in~\eqref{eq:JointCC} reads
	\begin{align}
		\nonumber c_{k_1,k_2}{=}\langle{:}
			\binom{N}{k_1}\left(e^{-\frac{\eta}{2N}\hat n(-\beta)-\nu}\right)^{N-k_1}\left(\hat 1{-}e^{-\frac{\eta}{2N}\hat n(-\beta)-\nu}\right)^{k_1}\\
			{\times}\binom{N}{k_2}\left(e^{-\frac{\eta}{2N}\hat n(\beta)-\nu}\right)^{N-k_2}\left(\hat 1{-}e^{-\frac{\eta}{2N}\hat n(\beta)-\nu}\right)^{k_2}
		{:}\rangle
	\end{align}
	The theory of balanced homodyne detection with click detectors has been established in~\cite{SVA15}.
	Here, we will recall some of its features and complete the discussion of some aspects not considered previously.

	When working with photoelectric detectors in the balanced four-port scheme, one subtracts the photoelectric counts of the two detectors from one another.
	In the limit of strong LO, one measures the quadrature
	\begin{align}
		\hat  x(\varphi)=\hat a_{\rm SI} e^{-i\varphi}+\hat a_{\rm SI}^\dagger e^{i\varphi}
	\end{align}
	of the signal.
	Analogously, the click difference counts can be analyzed, which yields the moments of a nonlinear click-quadrature operator~\cite{SVA15},
	\begin{align}\label{B4PH_Eq_X}
		\langle{:}\hat X^m(\varphi){:}\rangle=&\langle{:}(\hat \pi_1-\hat \pi_2)^m{:}\rangle
		\\=&\sum_{j=0}^m\binom{m}{j}(-1)^{m-j}\langle{:}\hat \pi_1^{j}\hat \pi_2^{m-j}{:}\rangle,\nonumber
	\end{align}
	with
	\begin{align}\label{eq:BHDX}
	 	\hat X(\varphi)=2N{\textrm{e}}^{{-}\frac{\eta}{2N}|\beta|^2{-}\nu}&{:}e^{-\frac{\eta}{2N}\hat n}\sinh\left[\frac{\eta|\beta|}{2N}\hat x(\varphi)\right]{:}.
	\end{align}
	The quadrature $\hat x(\varphi)$ appears here in a hyperbolic sine function, which is a skew-symmetric, analytical function, $\sinh(x)=(e^{x}-e^{-x})/2=-\sinh(-x)$.
	It transforms the features we expect from a quadrature and thereby justifies the notion of a nonlinear quadrature.
	Consequently, a nonlinear squeezing condition can be given as
	\begin{align}\label{B4PH_Sq}
	 	0>\langle{:}[\Delta\hat X(\varphi)]^2{:}\rangle.
	\end{align}

	Moreover, additional phase-sensitive information can be obtained from the jointly measured click statistics, as it reveals some fundamental differences from the photoelectric detection model.
	In particular, the sum of click counts does, in contrast to its photoelectric counterpart, show phase-dependent behavior.
	The number of the sum of photoelectric counts is independent of the phase,
	\begin{align}
		(\eta\hat n_1+\nu)+(\eta\hat n_2+\nu)=\eta\hat a^\dagger_{\rm SI}\hat a_{\rm SI}+\eta|\beta|^2+2\nu.
	\end{align}
	The sum of click counts,
	\begin{align}
	 	\hat\pi_1{+}\hat\pi_2=2N-&2N{e}^{-\frac{\eta}{2N}|\beta|^2-\nu}{:}{e}^{-\frac{\eta}{2N}\hat n}\cosh\left[\frac{\eta|\beta|}{2N}\hat x(\varphi)\right]{:},
	\end{align}
	clearly shows a phase-dependent behavior.
	Thus, a nonclassicality criterion based on the variance of the sum can be formulated similarly to the case of the click difference.

	\begin{figure}[ht]
		\includegraphics[width=8.7cm]{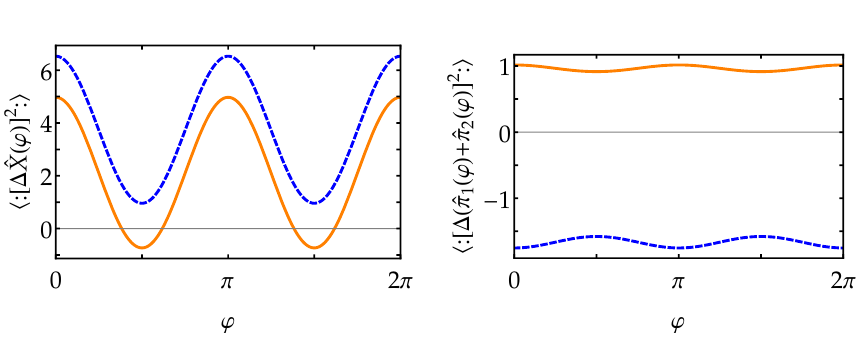}
	 	\caption{
	 		(Color online)
	 		The result of phase-dependent click statistics for a balanced homodyning scheme.
			The left plot shows the application of nonlinear squeezing criterion in Eq.~\eqref{B4PH_Sq}, and the right plot (scaled $\times10^2$) depicts the corresponding variance of the sum of clicks.
			The plotted states are $|\alpha_-\rangle$ (dashed line) and $|\alpha_+\rangle$ (solid line), where $\alpha=1$ and $|\beta|=4$.
	 		The detection parameters are $\eta=50\%$, $\nu=0$, $N=8$.
		}\label{B4PH_Fig_M_V}
	\end{figure}

	Examples of both criteria are shown for the even and odd coherent states~\eqref{eq:EvenOddState} in Fig.~\ref{B4PH_Fig_M_V}. 
	Nonlinear squeezing can be observed for the even coherent state, whereas a sub-shot-noise variance of the sum of clicks, $\langle{:}[\Delta(\hat\pi_1{+}\hat\pi_2)]^2{:}\rangle<0$, is visible for the odd state.
	Moreover, it is worth mentioning that we are not limited to weak or strong LOs.

\section{Imperfections}\label{sec:noise}
	Let us now discuss the influence of imperfections, which is crucial for the verification of nonclassicality~\cite{KLL03,BPP98,A87,STG08}.
	From the nonlinear structure of the mean click operators, e.g., Eqs.~\eqref{pi_plain} and~\eqref{4PH_Eq_pi1_Dn}, it is directly evident that click detectors respond differently to attenuations than photoelectric detectors.
	Moreover, the impact of a mode mismatch between the LO and the SI has not been studied so far.
	We will therefore study these realistic perturbations for the determination of nonclassicality while restricting ourselves to the fundamental case of unbalanced homodyne detection.

\subsection{Imperfections of the click-counting detectors}
	One imperfection is due to the dark count rate $\nu$ [see Eqs.~\eqref{eq:ClickStatistics} and~\eqref{eq:unbalancedCC}].
	That is, some clicks are recorded even if there was no SI or LO field.
	Let us consider the nonclassicality criteria, i.e., moments of the form~\eqref{4PH_Eq_pi1_Dn}.
	Comparing the case $\nu=0$ and a nonzero $\nu$, we can decompose the matrix of moments as
	\begin{align}
		&\left.\boldsymbol M\right|_{\nu>0}=\left(\langle{:}[N-\hat\pi(\varphi)]^{m+m'}{:}\rangle\right)_{m,m'=0}^{\lfloor N/2\rfloor}\nonumber
		\\=&\left(e^{-\nu[m+m']}\langle{:}[Ne^{-\frac{\eta_t}{N}\hat n(\gamma)}]^{m+m'}{:}\rangle\right)_{m,m'=0}^{\lfloor N/2\rfloor}\label{eq:noisecontribution}
		\\=&\boldsymbol T_{\nu}\left.\boldsymbol M\right|_{\nu=0}\boldsymbol T_{\nu},
		\text{ with } \boldsymbol T_{\nu}={\rm diag}\left(e^{-0\nu},\ldots,e^{-\lfloor \frac{N}{2}\rfloor \nu}\right).\nonumber
	\end{align}
	This transformation property between the cases with and without dark counts allows us to state the following:
	If nonclassicality can be detected for no dark counts, $\nu=0$, then nonclassicality can be detected for a finite dark count rate, $\nu>0$.
	Thus, the impact of dark counts, excluding the case $\nu\to\infty$, is not an issue for the verification of nonclassicality.
	It solely scales the actual values of the minors, e.g.,
	\begin{align}
		\left.\langle{:}[\Delta\hat\pi(\varphi)]^2{:}\rangle\right|_{\nu>0}=\left.\langle{:}[\Delta\hat\pi(\varphi)]^2{:}\rangle\right|_{\nu=0}\times e^{-2\nu}.
	\end{align}
	Thus, we can assume for our theoretical studies $\nu=0$.
	In experiments $\nu$ can also be estimated, e.g., $\nu\approx 0$~\cite{LFCPSREPW09} or $\nu\approx 0.5$~\cite{SBVHBAS15}.

	Such a simple treatment is not possible when considering a non unit quantum efficiency, $\eta<1$.
	This analysis has to be performed specifically for the desired target state.
	Such a test, in advance of an experiment, is helpful for estimating the overall efficiency one requires to infer nonclassicality.

\begin{figure}[ht]
	 	\includegraphics[width=6.3cm]{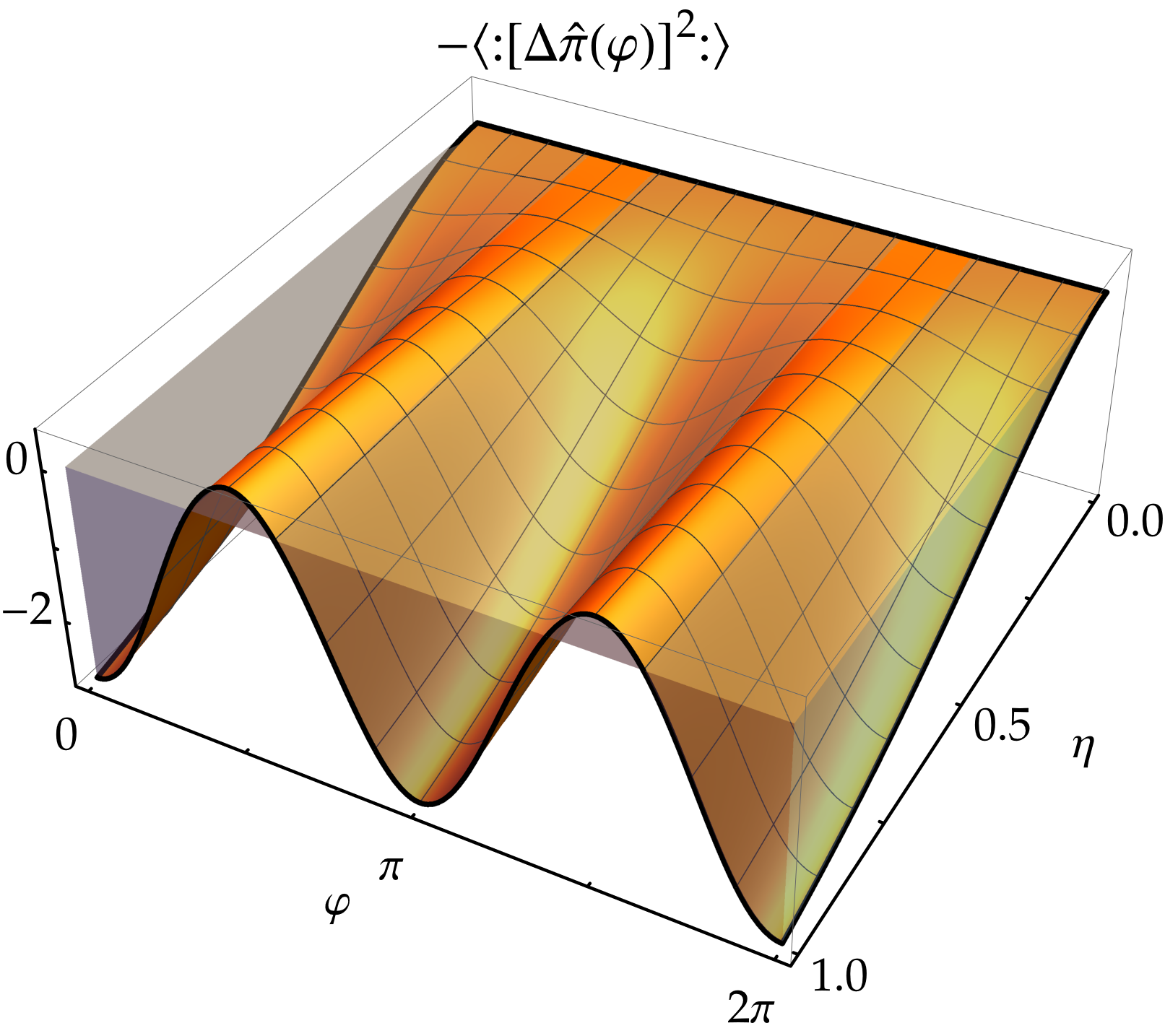}
	 	\caption{
		 	(Color online)
		 	The nonclassicality criterion~\eqref{4PH_Eq_pi_2nd_order_minor_criterion} is shown depending on the quantum efficiency $\eta$ and the phase $\varphi$.
		 	For better visibility, the negative variance, $-\langle{:}[\Delta\hat\pi(\varphi)]^2{:}\rangle$, is plotted.
		 	We consider an even coherent state $|\alpha_+\rangle$, where $\alpha=1$ and $\beta=4$,
		 	the click detector parameters are $\nu=0$ and $N=8$,
		 	and the beam splitter is characterized by $t=4/5$ and $r=3/5$.
		 }\label{Imp_Fig_pi1_V}
	\end{figure}

	Let us take a closer look at the even coherent state (see Fig.~\ref{Imp_Fig_pi1_V}).
	For this state the appearance of nonclassicality seems to be quite independent of the quantum efficiency.
	That is, the value of the negativity is diminished for lowered efficiency, but the negativity is present for the $\pi/2$ and $3\pi/2$ phases for any $\eta>0$.
	However, the significance, with which the nonclassicality might be verified in experiments, decreases.

\subsection{Saturation effects}
	The number $N$ of APDs is finite.
	It can be shown (cf.~\cite{SVA12_85,SVA15}) that we approach a true displaced number operator,
	\begin{align}
		\langle{:}\hat\pi^m(\varphi){:}\rangle
		=\langle{:}N^m(1{-}e^{-\frac{\eta_t}{N}\hat n(\gamma)})^m{:}\rangle
		\approx \langle{:}[\eta_t\hat n(\gamma)]^m{:}\rangle,
	\end{align}
	in the limit $N\to\infty$.
	Another version of the limit would be that the powers of photon numbers are comparably small, $\langle{:}\hat n^m(\gamma){:}\rangle\ll N^m$, for $m\in\mathbb N\setminus\{0\}$.

	Both approximations are typically not justified.
	Even worse, accepting these approximations would yield nonclassicality for classical coherent states~\cite{SVA12_109}.
	To counter these fake effects, all our nonclassicality criteria are formulated in terms of moments of the click statistics.
	This allows us to treat all ranges of intensities without approximations.
	This also means that saturation effects are included in our click-counting theory.
	Namely, the case $\langle{:}\hat n^m(\gamma){:}\rangle\gtrsim N^m$ is properly described.
	For $m=1$ this means that the mean photon number $\langle\hat n(\gamma)\rangle$ can be on the same order as or exceed the number of on-off diodes, $N$.
	
	Let us show that fake nonclassicality will not occur for high intensities.
	Thus, we may assume a quantum SI with many photons and a classical SI with the same intensity.
	Basically, the high-intensity limit of both cases yields that all APDs click at the same time, i.e.,
	\begin{align}
		c_k\approx\left\lbrace\begin{array}{ccl}
			0 & \text{ for } & k=0,\ldots, N-1,\\
			1 & \text{ for } & k=N.
		\end{array}\right.
	\end{align}
	Note that the same is true if the LO is strong.
	For the moments [see formula~\eqref{eq:samplingformula}] we get
	\begin{align}
		\langle{:} \hat\pi^m(\varphi){:}\rangle=N^m\sum_{k=m}^N\frac{\binom{k}{m}}{\binom{N}{m}}c_k\approx N^m,
	\end{align}
	which consistently yields a positive matrix of moments, $(N^{m+m'})_{m,m'}>0$, for the classical and nonclassical states.
	In conclusion, (i) too high intensities are not helpful for the determination of nonclassicality, and (ii) a proper detector description will not yield fake nonclassicality, even in case the of saturation.

\subsection{Imperfections due to the local oscillator}
	Imperfections in homodyne detection may also stem solely from the LO.
	Typical examples in realistic measurement scenarios are fluctuations of the LO itself or a mode mismatch between the LO and the SI fields.
	Let us discuss the implications for such sources of errors.
	Explicitly, the impact on $m$th order moments will be studied,
	\begin{align}\label{eq:no-noise}
		\langle {:}[N-\hat \pi(\varphi)]^m{:}\rangle=(Ne^{-\nu})^m\langle{:}e^{-\frac{m}{N}\eta_t(\hat a_{\rm SI}-\gamma)^\dagger(\hat a_{\rm SI}-\gamma)}{:}\rangle,
	\end{align}
	with $\gamma=-r\beta/t$ ($\varphi=\arg\gamma$) being a linear function of the amplitude of the LO state, $|\beta\rangle_{\rm LO}$.

	For small amplitudes some noise of the LO may occur, e.g., due to thermal fluctuations.
	For the balanced four-port homodyne detection, phase and amplitude noise have been studied in Ref.~\cite{SVA15}.
	Here, we will focus on perturbations in the unbalanced scenario.
	If the source of noise is a classical one, we may describe this effect by a classical probability distribution, $P_{\rm LO}(\gamma)\geq0$, of the (scaled) LO amplitude.
	Thus, we get a convolution of the moment with noise:
	\begin{align}\label{eq:LOnoise}
		&\langle {:}[N-\hat \pi(\varphi)]^m{:}\rangle_{P_{\rm LO}}\\
		=&\int d^2\gamma'\,P_{\rm LO}(\gamma-\gamma')(Ne^{-\nu})^m\langle{:}e^{-\frac{m}{N}\eta_t(\hat a_{\rm SI}-\gamma')^\dagger(\hat a_{\rm SI}-\gamma')}{:}\rangle.\nonumber
	\end{align}
	Considering only thermal fluctuations, we can suppose that the LO is a displaced thermal state, i.e., $P_{\rm LO}(\gamma')=\exp(-|\gamma'|^2/\bar n)/(\pi\bar n)$.
	This allows us to compute the Gaussian integral~\eqref{eq:LOnoise} as
	\begin{align}\label{eq:thermal-noise}
		&\int d^2\gamma'\,P_{\rm LO}(\gamma-\gamma')\langle{:}e^{-\lambda(\hat a_{\rm SI}-\gamma')^\dagger(\hat a_{\rm SI}-\gamma')}{:}\rangle\\
		=&\frac{1}{1+\lambda\bar n}\langle{:}\exp\left(-\frac{\lambda}{1+\bar n\lambda}\hat n(\gamma)\right){:}\rangle,\nonumber
	\end{align}
	where $\lambda=\eta_tm/N\in[0,1]$.
	Note that the prefactor of the initial exponent $\lambda$ in Eq.~\eqref{eq:no-noise} is reduced to $\lambda/(1+\bar n\lambda)$ in Eq.~\eqref{eq:thermal-noise}.
	This can be regarded as a diminished quantum efficiency.
	The analytical formula~\eqref{eq:thermal-noise} may be used to estimate the impact of thermal LO fluctuations on the nonclassicality probes.

\begin{figure}[ht]
	 	\includegraphics[width=6.3cm]{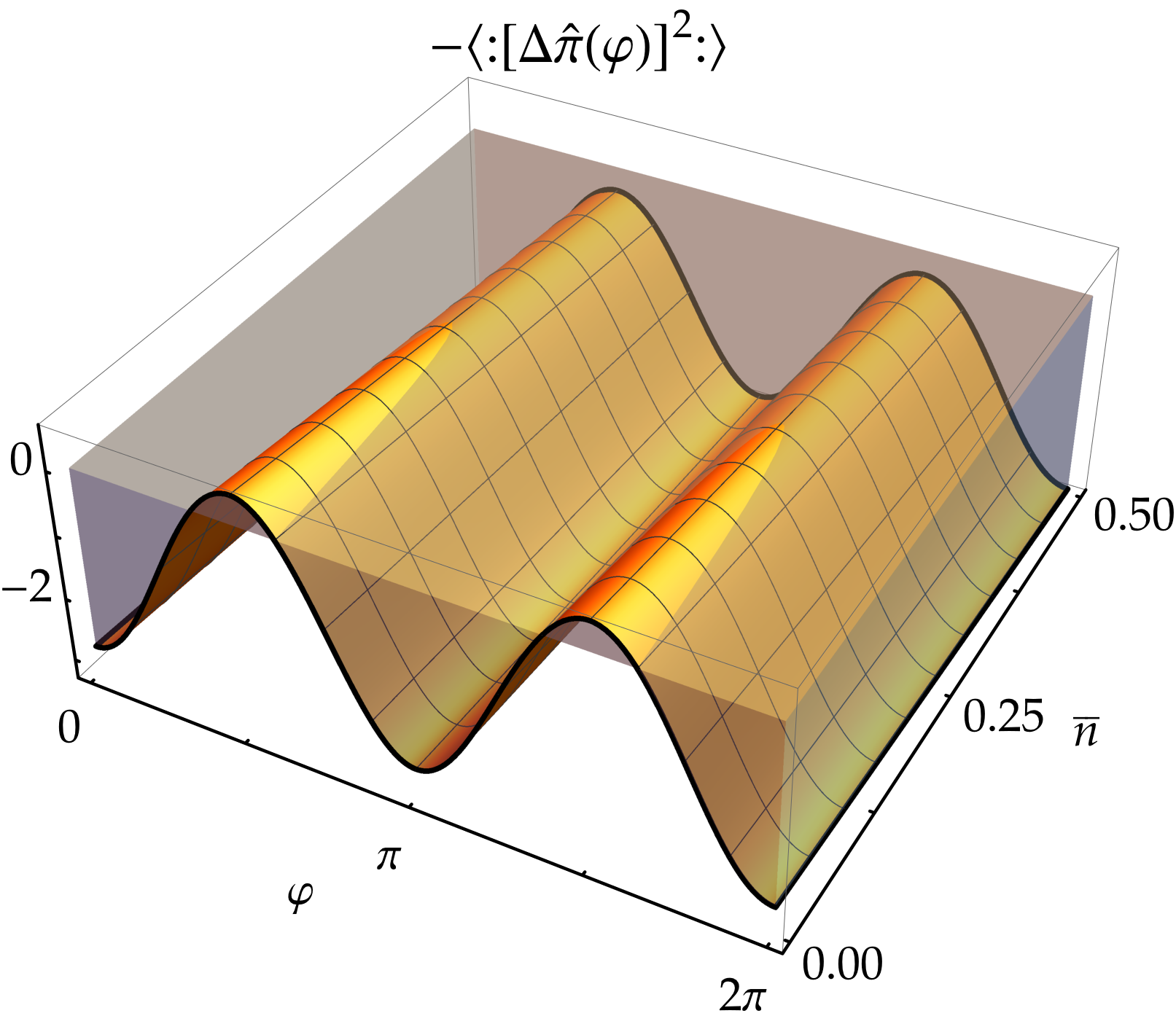}
	 	\caption{
			(Color online)
		 	The nonclassicality criterion~\eqref{4PH_Eq_pi_2nd_order_minor_criterion} is shown for the even coherent state $|\alpha_+\rangle$ superimposed with a LO,
		 	where $\alpha=1$ and $\beta=4$.
		 	The negative normally ordered variance of $\hat\pi(\varphi)$ is analyzed depending on the thermal photon number $\bar n$ of the LO and the phase $\varphi$.
		 	The detection and beam splitter parameters are $\eta=1$, $\nu=0$, $N=8$, $t=4/5$, and $r=3/5$.
		 	}\label{Fig_imp_LO}
	\end{figure}

	An example of the impact of a thermal LO on the measurement of a nonclassicality probe can be seen in Fig.~\ref{Fig_imp_LO}.
	For the even coherent state the nonclassicality cannot be determined for strong fluctuations of the LO.
	Namely, in the studied scenario the LO is much stronger than the SI, $|\alpha|^2\ll|\beta|^2$, which means that the click detector mainly detects the noisy LO.
	Thus, experiments should avoid such intensity relations, or the LO fluctuations have to be minimized.

	A second effect, which is due to the LO, is a mode mismatch.
	This imperfection occurs whenever the modes of the LO and the SI field do not have a perfect overlap when combining them on a beam splitter (see Fig.~\ref{4PH_scheme}).
	The rigorous derivation of the resulting modifications based on the spectral response of the detector can be found in Appendix~\ref{App:ModeMM}.

	Here, let us discuss solely the results.
	The moments change to
	\begin{align}
		\nonumber &\langle {:}[N-\hat \pi(\varphi)]^m{:}\rangle_{\rm mismatch}
		\\=&(Ne^{-(\nu+\tilde \nu)})^m\langle{:}e^{-\frac{m}{N}\eta_t(\hat a_{\rm SI}-\gamma)^\dagger(\hat a_{\rm SI}-\gamma)}{:}\rangle,
	\end{align}
	introducing an additional noise term $\tilde \nu$.
	In particular, this noise contribution is proportional to the intensity of the LO beam, 
	\begin{align}
		\tilde\nu\propto|\gamma|^2
	\end{align}
	(see Appendix~\ref{App:ModeMM}).
	As we have seen before [Eq.~\eqref{eq:noisecontribution}], such a noise contribution will not affect the nonclassical properties of the matrix of click-counting moments.
	This situation changes in combination with LO fluctuations, since it will contribute to the convolution~\eqref{eq:LOnoise}.
	Therein, $\nu$ has to be replaced by $\nu+\tilde \nu$.

\section{Multi-port homodyne detection}\label{sec:4}
	\begin{figure}[ht]
	 	\includegraphics[width=5.64cm]{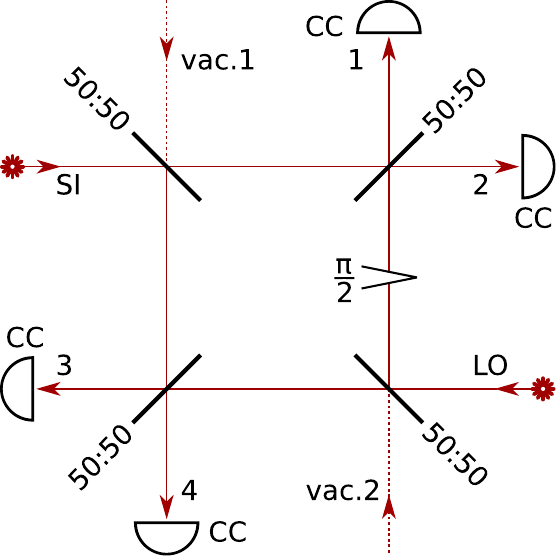}
	 	\caption{
	 		(Color online)
	 		Balanced eight-port homodyning scheme consisting of four $50{:}50$ beam splitter and a  $\pi/2$ phase shifter.
			The SI and LO are each fed into an input port, whereas two input ports, vac.1 and vac.2, remain unused.
			The outgoing fields are detected with click-counting detectors.
			Nonlinear click quadratures can be subsequently obtained by subtracting the click counts of the detectors in positions 1 and 2 as well as positions 3 and 4.
		}\label{B8PH_Fig_scheme}
	\end{figure}

	After this detailed consideration of imperfections, let us proceed by generalizing the formalism of four-port detection with click-counting detectors to prominent balanced multi-port homodyning schemes.
	A standard balanced setup is the well-known eight-port scheme, shown in Fig.~\ref{B8PH_Fig_scheme} for four click detectors.
	Its input-output relation can be written as~\cite{VW06}
	\begin{align}
		\begin{pmatrix}
	 		\hat a_1\\\hat a_2\\\hat a_3\\\hat a_4
	 	\end{pmatrix}
	 	=\frac{1}{\sqrt{4}}
	 	\begin{pmatrix}
	 		-1 & 1 & 1 & -i \\
	 		1 & 1 & -1 & -i \\ 
	 		1 & i & 1 & -1 \\
	 		1 & -i & 1 & 1
	 	\end{pmatrix}
	 	\begin{pmatrix}
			\hat a_{\textrm{SI}}\\\hat a_{\textrm{LO}}\\\hat a_\text{vac.1}\\\hat a_\text{vac.2}
	 	\end{pmatrix}.
	\end{align}
	Again, we assume that all click detectors have the same characteristics.
	Thus, generalizing the previously introduced approaches, we get
	\begin{align}\label{eq:8PMoments}
		&\langle{:}(\hat\pi_1-\hat\pi_2)^k(\hat\pi_3-\hat\pi_4)^l{:}\rangle
		\\\nonumber = & \langle{:}
			\left(2Ne^{-[\frac{\eta}{4N}|\beta|^2+\nu]}\right)^{k}
			e^{-k\frac{\eta}{4N}\hat n}
			\sinh^k\left[\frac{\eta|\beta|}{4N}\hat x(\varphi)\right]
		\\\nonumber & \times
			\left(2Ne^{-[\frac{\eta}{4N}|\beta|^2+\nu]}\right)^{l}
			e^{-l\frac{\eta}{4N}\hat n}
			\sinh^l\left[\frac{\eta|\beta|}{4N}\hat x(\varphi+\frac{\pi}{2})\right]{:}\rangle,
	\end{align}
	with $k,l=0,\ldots,N$.
	The conjugate momentum to $\hat x(\varphi)$ is $\hat p(\varphi)=\hat x(\varphi+\frac{\pi}{2})$.
	This can be adopted to define the nonlinear momentum operator,
	\begin{align}
		\hat P(\varphi)=&2Ne^{-[\frac{\eta}{4N}|\beta|^2+\nu]}
			{:}e^{-\frac{\eta}{4N}\hat n}
			\sinh\left[\frac{\eta|\beta|}{4N}\hat p(\varphi)\right]{:},
	\end{align}
	for the nonlinear quadrature operator, here in the form
	\begin{align}
		\hat X(\varphi)=&2Ne^{-[\frac{\eta}{4N}|\beta|^2+\nu]}
			{:}e^{-\frac{\eta}{4N}\hat n}
			\sinh\left[\frac{\eta|\beta|}{4N}\hat x(\varphi)\right]{:}.
		\label{B8PH_Eq_X}
	\end{align}

	Let us point out that $\hat X(\varphi)$ in Eq.~\eqref{eq:BHDX} for the four-port scheme includes terms which scale with $\eta/(2N)$, whereas for the eight-port scheme we have a scaling with $\eta/(4N)$ [see Eq.~\eqref{B8PH_Eq_X}].
	This deficiency could be corrected by taking half the numbers of APDs for the eight-port homodyne detection.
	Using photoelectric detectors, such a scaling also occurs~\cite{VW06,ZVW96}.
	There, however, the correction requires us to have a doubled quantum efficiency in the eight-port scheme instead, which is a quite demanding task.
	In click detection, the total efficiency of the detector system, consisting of $N$ APDs, each having a quantum efficiency $\eta$, is given by the fraction $\eta/N$.
	This allows us to modify $N$ or $\eta$ for manipulating the overall efficiency.
	This relates to the findings in Ref.~\cite{LSV15}, where it has been demonstrated that it can be advantageous to have fewer on-off detectors in some scenarios.

	Finally, we can formulate nonclassicality in terms of variances, $0>\langle{:}[\Delta\hat X(\varphi)]^2{:}\rangle$ or $0>\langle{:}[\Delta\hat P(\varphi)]^2{:}\rangle$, or we can uncover nonclassical correlations between click position and momentum via
	\begin{align}\label{B8PH_Covar}
		0>\langle{:}[\Delta\hat X(\varphi)]^2{:}\rangle\langle{:}[\Delta\hat P(\varphi)]^2{:}\rangle-\langle{:}\Delta\hat X(\varphi)\Delta\hat P(\varphi){:}\rangle^2.
	\end{align}
	In this form, the nonclassicality condition relates to a violation of a normally ordered version of the Schr\"odinger-Robertson uncertainty relation~\cite{CommentSRUncertainty}.
	Let us stress again that not only can moment-based nonclassicality criteria be constructed from the second-order difference moments, but more general criteria may be considered as well.
	For instance, the variance of the sum of click events from different detectors could be used to certify the quantum character of the odd coherent state.

\begin{figure}[ht]
		 \includegraphics[width=8.7cm]{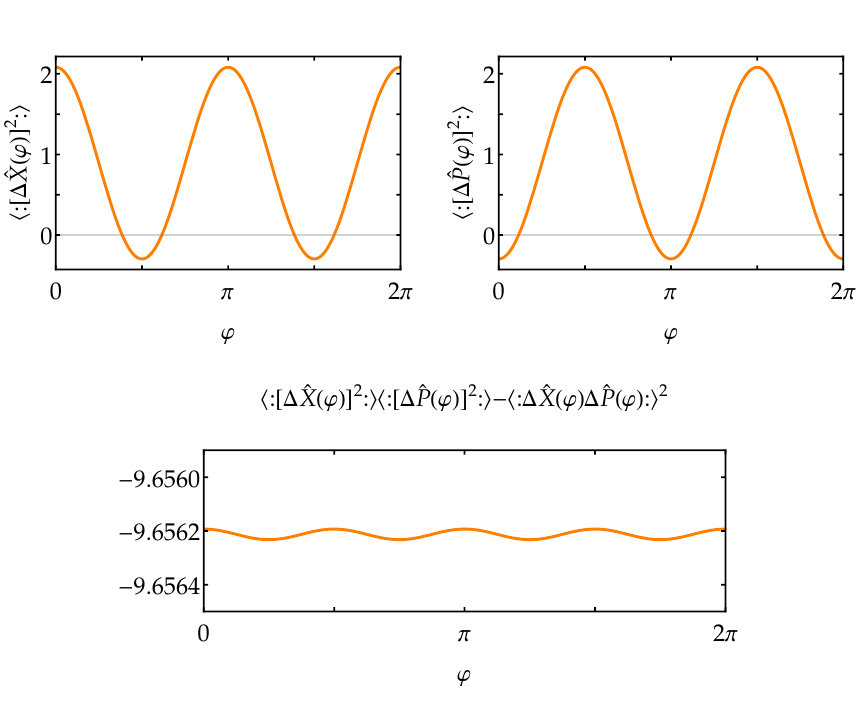}
		\caption{
			(Color online)
			Both click quadratures variances [top left: $\hat X(\varphi)$, top right: $\hat P(\varphi)$] and the covariance in Eq.~\eqref{B8PH_Covar} (bottom, scaled $\times10^{3}$) are shown as a function of the phase $\varphi$ for the even coherent state $|\alpha_+\rangle$ with $\alpha=1$.
			As considered before, the detection parameters are $\eta=50\%$, $\nu=0$, $N=8$, and the LO amplitude is $|\beta|=4$.
	 	}\label{B8PH_Fig_X1_X2_M_V}
	\end{figure}

	Examples for the single quadrature variances and the covariance are given in Fig.~\ref{B8PH_Fig_X1_X2_M_V}.
	Nonclassicality is determined for the even coherent state.
	Here, it is worth pointing out that the evaluated condition~\eqref{B8PH_Covar} includes very small oscillations, and it is negative for any phase $\varphi$.
	The nonlinear position and momentum variances are negative only for small phase intervals.

	In general, the method derived here applies to all kinds of multi-port homodyning schemes.
	For example, a six-port scheme with photoelectric detectors was proposed in Ref.~\cite{ZVW96} and could be similarly formulated with click-counting techniques.
	Since the six-port scenario includes only one vacuum input, it yields smaller attenuations than the eight-port scheme with two vacuum inputs discussed here (see Fig.~\ref{B8PH_Fig_scheme}).

\section{Multi-Mode Measurements}\label{sec:5}
\begin{figure}[ht]
	 	\includegraphics[width=6.225cm]{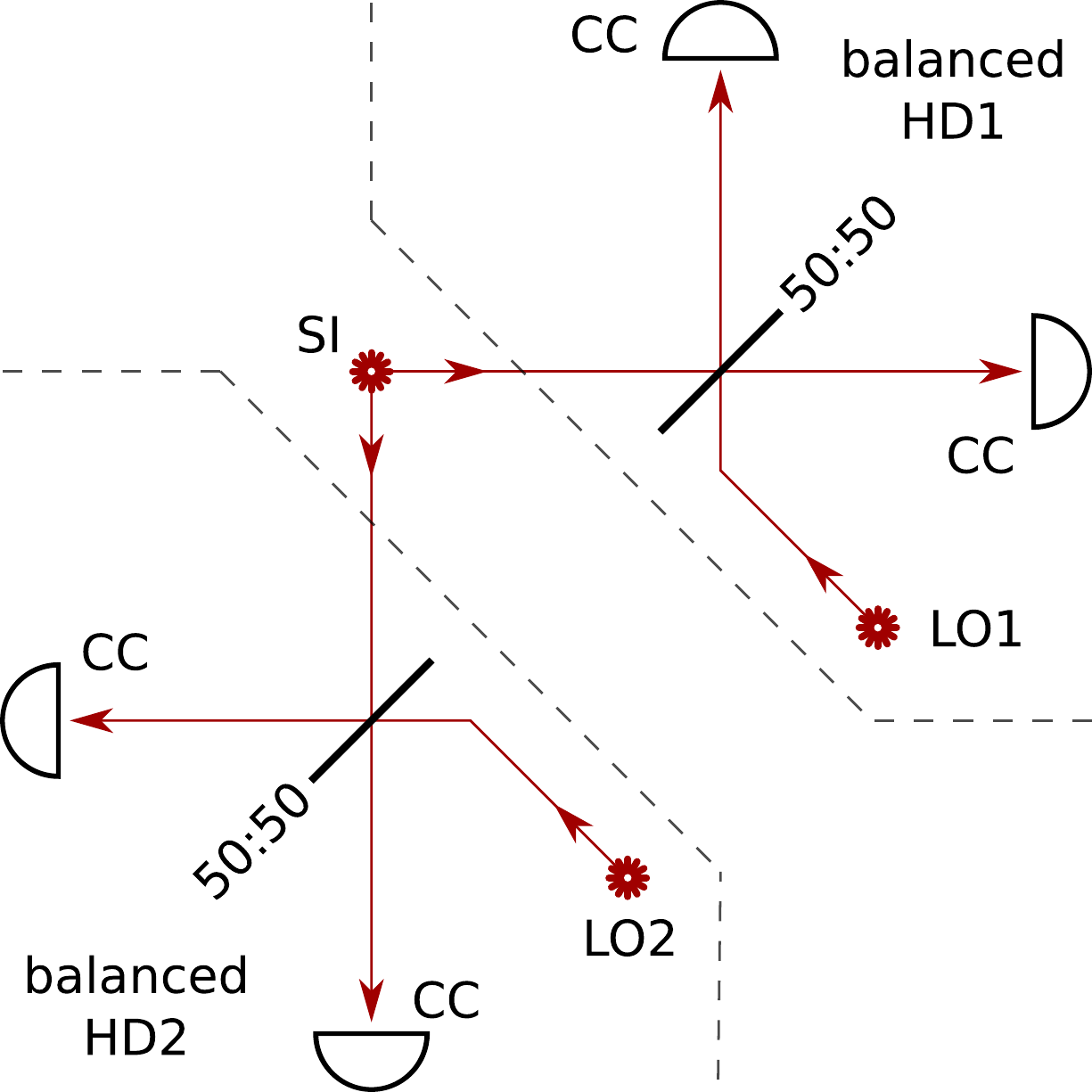}
	 	\caption{
	 		(Color online)
	 		A two-mode correlation measurement scheme consisting of four click counters.
			Each of the two SI and LO beams are fed into a balanced homodyne detection (``HD1'' and ``HD2'') four-port scheme.
			The individual, nonlinear click quadratures of the two modes, $\hat X_i(\varphi_i)$ for $i=1,2$, are subsequently obtained by subtracting the click counts in each setting.
	 		}
	 	\label{Fig_B2MH_scheme}
	\end{figure}

	So far, we have considered only a single SI field.
	Here, we outline the multimode scenario of phase-sensitive click detectors at different positions, which is especially interesting for determining quantum correlations between beams.
	For simplicity, we restrict our discussion to two-mode balanced homodyne detectors as one example for a correlation measurement (see Fig.~\ref{Fig_B2MH_scheme}).
	A generalization to multi-port schemes with more than two spatial modes or with other homodyning schemes is straightforward.

	In Fig.~\ref{Fig_B2MH_scheme}, each of the signal modes $\hat a_{\rm SI.1}$ and $\hat a_{\rm SI.2}$ is fed into a balanced homodyne.
	Therein they are superimposed with the corresponding LO modes of coherent states $|\beta_i\rangle$, with $i=1,2$.
	The click difference counts in each balanced detectors yield two nonlinear quadratures $\hat X_1(\varphi_1)$ and $\hat X_2(\varphi_2)$ (see Sec.~\ref{sec:3:b}).
	Single-mode nonclassicality for the SI in mode $i=1$ or $i=2$ is verified if the corresponding normally ordered variance is negative,
	\begin{align}\label{B2MH_Eq_X1_X2}
	 	0>\langle:[\Delta \hat X_i(\varphi_i)]^2:\rangle.
	\end{align}
	Additionally, nonclassical two-mode correlation between the two signals can be inferred from a negative covariance:
	\begin{align}\label{B2MH_Eq_C12}
	 	0>&\langle{:}[\Delta \hat X_1(\varphi_1)]^2{:}\rangle\langle{:}[\Delta \hat X_2(\varphi_2)]^2{:}\rangle
	 	\\&-\langle{:}\Delta \hat X_1(\varphi_1)\Delta \hat X_2(\varphi_2){:}\rangle^2,\nonumber 
	\end{align}
	which yields a nonlinear two-mode squeezing.
	It is also worth mentioning that a quantum correlation between the SIs in terms of a sum of the local balanced homodyne click counting devices can be formulated, as  done for the single-mode case in Sec.~\ref{sec:3:b}.

\begin{figure}[ht]
		\includegraphics[width=8.7cm]{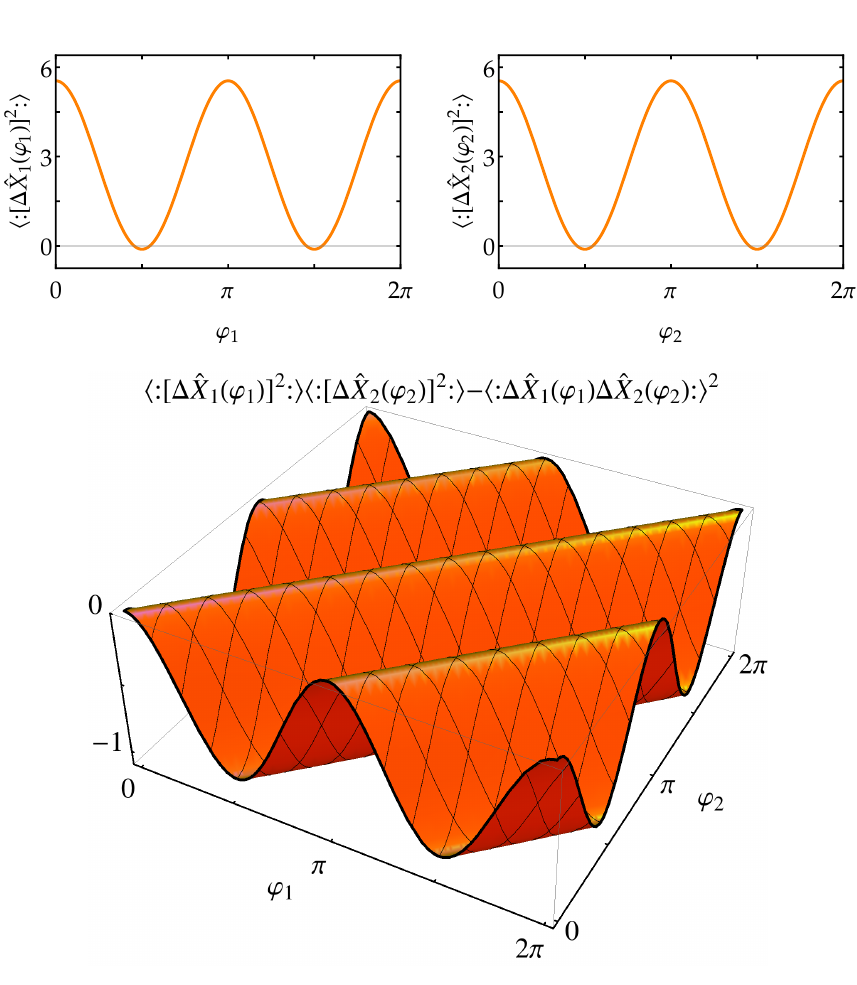}
		\caption{
			(Color online)
			The click quadrature variances~\eqref{B2MH_Eq_X1_X2} (top left: $i=1$, top right: $i=2$) and the covariance in Eq.~\eqref{B2MH_Eq_C12} (bottom, scaled $\times10^2$) are shown as a function of the phases $\varphi_i$, $i=1,2$.
			Nonclassicality of the two-mode even coherent state $|\alpha_+\rangle$, with $\alpha=1$, is revealed for click-counting detection parameters $\eta=50\%$, $\nu=0$, $N=8$, and the LO amplitudes are $|\beta_1|=|\beta_2|=4$.
		}\label{B8PH_Fig_X1_X2}
	\end{figure}

	Let us apply this approach.
	The natural extensions of the single-mode even and odd coherent states in Eq.~\eqref{eq:EvenOddState} are the two-mode odd and even coherent states,
	\begin{align}
	 	|\alpha^{(2)}_\pm\rangle=\frac{|\alpha,\alpha\rangle\pm|-\alpha,-\alpha\rangle}{\sqrt{2[1\pm\exp(-4|\alpha|^2)]}}.
	\end{align}
	The results for the second-order criteria are illustrated in Fig.~\ref{B8PH_Fig_X1_X2}.
	For the given parameter range, the single-mode variances of the two-mode even coherent state are negative for a small neighborhood of $\pi/2$ and $3\pi/2$.
	In contrast, the nonclassical covariance between the modes is negative for almost all phase values even for a detection efficiency of $50\%$.

	Let us remark that multi-time, multi-detector correlation measurements as reported in Ref.~\cite{SV06} can be achieved with a similar approach.
	The multi-time click-counting theory was established~\cite{SVA13} under the constraint that each APD produces not more than one click in a given measurement interval.
	The corresponding nonclassicality conditions, e.g., for the click counterpart of the photon antibunching effect~\cite{KM76,WC76,KDM77}, have also been formulated in terms of normally and time-ordered matrices of click-counting moments.

\section{Summary and Conclusions}\label{sec:6}
	In summary, we have formulated a theoretical model for the implementation of click detectors in phase-sensitive homodyne measurement schemes.
	Based on such setups and employing the click-counting theory, we studied the verification of nonclassical light for realistic detection processes.
	Since our results have been expressed in terms of analytical formulas, they may be helpful for predicting the experimental results and for formulating bounds to imperfections in realized setups.
	In addition, the moments for the considered nonclassicality probes can be directly derived from measured click statistics, making our findings easily accessible for experimental implementations.

	We formulated the four-port homodyning including the balanced and unbalanced detection scenarios.
	Our approach in terms of click-counting detectors was compared with the traditional scenario using photoelectric detection models.
	Perturbations stemming from the imperfect click-counting detectors, e.g., efficiencies and saturation effects, or the impurities of the local oscillator have been studied.
	Moreover, the influence of a mode mismatch between the signal and the local oscillator has been shown to result in an additional dark count rate which is proportional to the intensity of the local oscillator.
	In the case of multi-port homodyning, we identified a nonclassicality condition in the form of an uncertainty relation between the nonlinear position and momentum operators.
	In the same fashion, quantum correlations between multiple signal fields have been studied.
	It is worth mentioning that we focused our consideration of nonclassicality probes on second-order criteria.
	The extension to higher-order moments was also discussed.

	We can conclude that click detectors are capable of determining nonclassicality in various phase-sensitive optical measurement scenarios.
	This supports the assumption that click-counting devices can be employed whenever photoelectric detectors are not available.
	Hence, our technique offers a useful set of tools for current and future experiments.
	It may also be the starting point for the development of a tomographic state reconstruction approach with click-counting detectors, which requires further studies.

\section*{Acknowledgments}
	This work was supported by the Deutsche Forschungsgemeinschaft through SFB~652.

\appendix

\section{Matrix of Moments Expansion}\label{App:MOMexpansion}
	Nonclassicality conditions can be written in terms of matrices of moments.
	The typically considered matrix of normally ordered click-counting moments reads
	\begin{align}
		\boldsymbol M=\left(\langle{:}\hat \pi^{m+m'}{:}\rangle\right)_{m,m'}.
	\end{align}
	Note that $\langle{:}\hat\pi{:}\rangle{\geq}0$.
	Let us formulate two equivalent representations of matrices: (i) in terms of powers of $N-\hat\pi$, $\boldsymbol M'{=}\big(\langle{:}(N{-}\hat\pi)^{m+m'}{:}\rangle\big)_{m,m'}$, or (ii) in terms of centered moments $\Delta\hat\pi$, $\boldsymbol M''{=}\big(\langle{:}(\Delta\hat\pi)^{m+m'}{:}\rangle\big)_{m,m'}$.
	All these notations may be rewritten in the general form
	\begin{align}
		\boldsymbol M(x,y)=&\left(\langle{:}(x\hat 1+y\hat\pi)^{m+m'}{:}\rangle\right)_{m,m'},
	\end{align}
	with $x,y\in\mathbb R$ and $y\neq0$.
	Thus, we have $\boldsymbol M'{=}\boldsymbol M(N,{-}1)$ and $\boldsymbol M''{=}\boldsymbol M({-}\langle{:}\hat\pi{:}\rangle,1)$.
	We will prove that the non-negativity is preserved for all real parameters $x$ and $y\neq0$; that is, we claim that
	\begin{align}\label{Eq:MOMclaim}
		\boldsymbol M=\boldsymbol M(0,1)\geq0 \text{ $\Leftrightarrow$ } \boldsymbol M(x,y)\geq0.
	\end{align}

	To do so, let us expand an arbitrary matrix element
	\begin{align}\label{Eq:MoMexpansion}
		&\langle{:}(x\hat 1+y\hat\pi)^{m+m'}{:}\rangle
		\\\nonumber =&\sum_{k=0}^{m}\sum_{k'=0}^{m'}\binom{m}{k}\binom{m'}{k'}x^{m-k}x^{m'-k'}y^{k}y^{k'}\langle{:}\hat\pi^{k+k'}{:}\rangle.
	\end{align}
	Now, we can define a matrix $\boldsymbol T(x,y)=(t_{m,k})_{m,k}$ with
	\begin{align}
		t_{m,k}=\left\lbrace\begin{array}{ccc}
			0 & \text{ for } & k>m, \\
			\binom{m}{k}x^{m-k}y^{k} & \text{ for } & k\leq m.
		\end{array}\right.
	\end{align}
	The matrix $\boldsymbol T(x,y)$ is an upper triangular matrix with non zero diagonal elements, $t_{m,m}=y^m\neq0$.
	Therefore, $\boldsymbol T(x,y)$ is invertible.
	In addition, the expansion in Eq.~\eqref{Eq:MoMexpansion} proves the transformation
	\begin{align}\label{Eq:MoMtrafo}
		\boldsymbol M(x,y)=\boldsymbol T(x,y)\,\boldsymbol M(0,1)\,\boldsymbol T^{\dagger}(x,y).
	\end{align}
	Due to this transformation property and because $\boldsymbol T(x,y)^{-1}$ exists, the claim~\eqref{Eq:MOMclaim} holds true.

\section{Mode Mismatch}\label{App:ModeMM}
	We will derive the description of a mode mismatch between LO and SI for click detectors.
	This approach is based on multimode detection with spectral response functions (see~\cite{VW06,SVA13}).
	A single APD is properly characterized by two operators for the click, $\hat P_{\rm on}=\hat 1-{:}\exp[-\hat \Gamma]{:}$, and no-click event, $\hat P_{\rm off}={:}\exp[-\hat \Gamma]{:}$,
	where $\hat \Gamma$ is the detector response.
	Restricting ourselves to spectral properties (spatial and polarization degrees of freedom can be treated similarly) we have
	\begin{align}
		\hat\Gamma=\int d\omega\,G(\omega)\hat a^\dagger(\omega)\hat a(\omega),
	\end{align}
	with $G(\omega)\geq0$ being the so-called spectral response function of a broad band detector~\cite{VW06}.
	The annihilation operator for a frequency $\omega$ at the detector is $\hat a(\omega)$.
	For simplicity, we assume a vanishing dark count rate for the time being.

	It has been derived in Ref.~\cite{SVA13} that the click statistics for multimode fields may be expanded in terms of expectations values
	\begin{align}\label{Eq:MMMdesiredIntegral1}
		I(\lambda)=\langle {:}\exp[-\lambda\hat\Gamma]{:}\rangle,
		\text{ with }\lambda\in\{0/N,\ldots,N/N\}.
	\end{align}
	Moreover, the semi-classical representation in terms of the Glauber-Sudarshan representation of the SI field, $\hat\rho_{\rm SI}=\int d^2\alpha\, P(\alpha)  |\alpha\rangle\langle\alpha|$, allows one to consider coherent SI first, $|\alpha\rangle=|\alpha\rangle_{\rm SI}$, and generalize the considerations to arbitrary states by integrating the result with the $P$~function.
	The spectral decomposition of the LO and SI is
	\begin{align}
		\hat a_{i}=\int d\omega f_{i}(\omega)\hat a_{i}(\omega),
		\text{ for }i\in\{{\rm LO},{\rm SI}\}
	\end{align}
	and $\int d\omega |f_{i}(\omega)|^2=1$.
	Thus, the coherent SI and LO states can be written as
	\begin{align}
		|\alpha\rangle_{\rm SI}=\bigotimes_{\omega}|f_{\rm SI}(\omega)\alpha\rangle_{\omega}
		\text{ and }
		|\beta\rangle_{\rm LO}=\bigotimes_{\omega}|f_{\rm LO}(\omega)\beta\rangle_{\omega},
	\end{align}
	respectively.
	Applying a beam-splitter transformation to the spectral modes [$1=|t(\omega)|^2+|r(\omega)|^2$], we have for one output port of the beam splitter
	\begin{align}
		\hat a(\omega)=t(\omega)\hat a_{\rm SI}(\omega)+r(\omega)\hat a_{\rm LO}(\omega).
	\end{align}
	Hence, the desired expectation value~\eqref{Eq:MMMdesiredIntegral1} reads
	\begin{align}\label{Eq:MMMdesiredIntegral2}
		I(\lambda)=\exp\left[-\lambda (tf_{\rm SI}\alpha+rf_{\rm LO}\beta,tf_{\rm SI}\alpha+rf_{\rm LO}\beta)\right],
	\end{align}
	using the inner product
	\begin{align}
		(a,b)=\int d\omega\,G(\omega) a^\ast(\omega)b(\omega).
	\end{align}

	The derived expression~\eqref{Eq:MMMdesiredIntegral2} can be further rewritten as
	\begin{align}\label{Eq:MMMdesiredIntegral3}
		&I(\lambda)=\exp\left[-\lambda(tf_{\rm SI},tf_{\rm SI})\left|\alpha+\tfrac{(tf_{\rm SI},rf_{\rm LO})}{(tf_{\rm SI},tf_{\rm SI})}\beta\right|^2\right]
		\\&\times\exp\left[-\lambda\tfrac{
			(tf_{\rm SI},tf_{\rm SI})(rf_{\rm LO},rf_{\rm LO})-\left|(tf_{\rm SI},rf_{\rm LO})\right|^2
		}{(tf_{\rm SI},tf_{\rm SI})}|\beta|^2\right].\nonumber
	\end{align}
	Now we can define the overall quantum efficiency $\eta_t{=}(tf_{\rm SI},tf_{\rm SI})$, the coherent displacement $\gamma=-[(tf_{\rm SI},rf_{\rm LO})/(tf_{\rm SI},tf_{\rm SI})]\beta$, and a rate for the mode mismatch
	\begin{align}\label{Eq:MMMrate}
		\tilde \nu=\frac{
			(tf_{\rm SI},tf_{\rm SI})(rf_{\rm LO},rf_{\rm LO})-\left|(tf_{\rm SI},rf_{\rm LO})\right|^2
		}{(tf_{\rm SI},tf_{\rm SI})}|\beta|^2.
	\end{align}
	Note that Cauchy-Schwartz inequality implies $\tilde\nu\geq0$.
	Using the $P$~representation and the normal-ordering prescription, we get our desired quantity~\eqref{Eq:MMMdesiredIntegral1} for arbitrary SI states $\hat\rho_{\rm SI}$ from Eq.~\eqref{Eq:MMMdesiredIntegral3} as
	\begin{align}
		I(\lambda)=\langle{:}\exp\left[-\lambda\eta_t(\hat a_{\rm SI}-\gamma)^\dagger(\hat a_{\rm SI}-\gamma)-\lambda\tilde\nu \right]{:}\rangle.
	\end{align}
	In comparison to a perfect mode matching ($tf_{\rm SI}=rf_{\rm LO}$ $\Rightarrow$ $\tilde \nu=0$), we have (i) the same structure of an exponential of a displaced photon number operator $\hat n(\gamma)$ [see Eq.~\eqref{4PH_Eq_pi1_Dn}] and (ii) an additional dark count rate $\tilde\nu$ which is proportional to the intensity of the LO [see Eq.~\eqref{Eq:MMMrate}], or, equivalently,
	\begin{align}
		\tilde\nu\propto|\gamma|^2.
	\end{align}



\begin{thebibliography}{99}
	\bibitem{MW95}
		L. Mandel and E. Wolf,
		\textit{Optical Coherence and Quantum Optics}
		(Cambridge University Press, Cambridge, 1995).
	\bibitem{VW06}
		W. Vogel and D.-G. Welsch,
		\textit{Quantum Optics},
		3rd ed. (Wiley-VCH, Weinheim, 2006).
	\bibitem{A13}
		G. S. Agarwal,
		\textit{Quantum Optics}
		(Cambridge University Press, Cambridge, 2013).
	\bibitem{KWM00}
		A. Kuzmich, I. A. Walmsley, and L. Mandel,
		Violation of Bell's Inequality by a Generalized Einstein-Podolsky-Rosen State Using Homodyne Detection,
		Phys. Rev. Lett. \textbf{85}, 1349 (2000).
	\bibitem{PDFEPW10}
		G. Puentes, A. Datta, A. Feito, J. Eisert, M. B. Plenio, and I. A. Walmsley,
		Entanglement quantification from incomplete measurements: applications using photon-number-resolving weak homodyne detectors,
		New J. Phys. \textbf{12}, 033042 (2010).
	\bibitem{PADLA10}
		W. N. Plick, P. M. Anisimov, J. P. Dowling, H. Lee, and G. S. Agarwal,
		Parity detection in quantum optical metrology without number-resolving detectors,
		New J. Phys. \textbf{12}, 113025 (2010).
	\bibitem{CIDE14}
		L. Cohen, D. Istrati, L. Dovrat, and H. S. Eisenberg,
		Super-resolved phase measurements at the shot noise limit by parity measurement,
		Opt. Express \textbf{22}, 11945 (2014).
	\bibitem{WVO99}
		D.-G. Welsch, W. Vogel, and T. Opatrn\'y,
		Homodyne detection and quantum-state reconstruction,
		Prog. Opt. \textbf{39}, 63 (1999).
	\bibitem{LR09}
		A. I. Lvovsky and M. G. Raymer,
		Continuous-variable optical quantum-state tomography,
		Rev. Mod. Phys. \textbf{81}, 299 (2009).
	\bibitem{SV11}
		A. A. Semenov and W. Vogel,
		Fake violations of the quantum Bell-parameter bound,
		Phys. Rev. A {\bf 83}, 032119 (2011).
	\bibitem{TG65}
		U. M. Titulaer and R. J. Glauber,
		Correlation Functions for Coherent Fields,
		Phys. Rev. \textbf{140}, B676 (1965).
	\bibitem{M86}
		L. Mandel,
		Non-Classical States of the Electromagnetic Field,
		Phys. Scr. \textbf{T12}, 34 (1986).
	\bibitem{G63}
		R. J. Glauber,
		Coherent and Incoherent States of the Radiation Field,
		Phys. Rev. \textbf{131}, 2766 (1963).
	\bibitem{S63}
		E. C. G. Sudarshan,
		Equivalence of Semiclassical and Quantum Mechanical Descriptions of Statistical Light Beams,
		Phys. Rev. Lett. \textbf{10}, 277 (1963).
	\bibitem{NFM91}
		J. W. Noh, A. Foug\`{e}res, and L. Mandel,
		Measurement of the quantum phase by photon counting,
		Phys. Rev. Lett. \textbf{67}, 1426 (1991).
	\bibitem{SBRF93}
		D. T. Smithey, M. Beck, M. G. Raymer, and A. Faridani,
		Measurement of the Wigner distribution and the density matrix of a light mode using optical homodyne tomography: Application to squeezed states and the vacuum,
		Phys. Rev. Lett. \textbf{70}, 1244 (1993).
	\bibitem{SBCRF93}
		D. T. Smithey, M. Beck, J. Cooper, M. G. Raymer, and A. Faridani,
		Complete experimental characterization of the quantum state of a light mode via the Wigner function and the density matrix: application to quantum phase distributions of vacuum and squeezed-vacuum states,
		Phys. Scr. \textbf{T48}, 35 (1993).
	\bibitem{FS93}
		M. Freyberger and W. Schleich,
		Photon counting, quantum phase, and phase-space distributions,
		Phys. Rev. A \textbf{47}, R30 (1993).
	\bibitem{VG93}
		W. Vogel and J. Grabow,
		Statistics of difference events in homodyne detection,
		Phys. Rev. A \textbf{47}, 4227 (1993).
	\bibitem{MBAR95}
		M. Munroe, D. Boggavarapu, M. E. Anderson, and M. G. Raymer,
		Photon-number statistics from the phase-averaged quadrature-field distribution: Theory and ultrafast measurement,
		Phys. Rev. A \textbf{52}, R924 (1995).
	\bibitem{DBJVDBW14}
		G. Donati, T. J. Bartley, X.-M. Jin, M.-D. Vidrighin, A. Datta, M. Barbieri, and I. A. Walmsley,
		Observing optical coherence across Fock layers with weak-field homodyne detectors,
		Nat. Commun. \textbf{5}, 5584 (2014).
	\bibitem{L05}
		U. Leonhardt,
		{\it Measuring the Quantum State of Light}
		(Cambridge University Press, Cambridge, 2005)
	\bibitem{KM76}
		H. J. Kimble and L. Mandel,
		Theory of resonance fluorescence,
		Phys. Rev. A \textbf{13}, 2123 (1976).
	\bibitem{WC76}
		H. J. Carmichael and D. F. Walls,
		A quantum-mechanical master equation treatment of the dynamical Stark effect,
		J. Phys. B: At. Mol. Phys. \textbf{9}, 1199 (1976).
	\bibitem{KDM77}
		H. J. Kimble, M. Dagenais, and L. Mandel,
		Photon Antibunching in Resonance Fluorescence,
		Phys. Rev. Lett. \textbf{39}, 691 (1977).
	\bibitem{EFMP11}
		M. D. Eisaman, J. Fan, A. Migdall, and S. V. Polyakov,
		Single-photon sources and detectors ,
		Rev Sci. Instrum. \textbf{82}, 071101 (2011).
	\bibitem{BC10}
		G. S. Buller and R. J. Collins,
		Single-photon generation and detection,
		Meas. Sci. Technol. \textbf{21}, 012002 (2010).
	\bibitem{ALBPMHM13}
		A. Allevi, M. Lamperti, M. Bondani, J. Pe\v{r}ina, Jr., V. Mich\'alek, O. Haderka, and R. Machulka,
		Characterizing the nonclassicality of mesoscopic optical twin-beam states,
		Phys. Rev. A \textbf{88}, 063807 (2013).
	\bibitem{SSTT13}
		H. Shibata, K. Shimizu, H. Takesue, and Y. Tokura,
		Superconducting Nanowire Single-Photon Detector with Ultralow Dark Count Rate Using Cold Optical Filters,
		Appl. Phys. Express \textbf{6}, 072801 (2013).
	\bibitem{MVSHLGVBSMN13}
		F. Marsili, V. B. Verma, J. A. Stern, S. Harrington, A. E. Lita, T. Gerrits, I. Vayshenker, B. Baek, M. D. Shaw, R. P. Mirin, and S. W. Nam,
		Detecting single infrared photons with 93\% system efficiency,
		Nat. Photon. \textbf{7}, 210 (2013).
	\bibitem{RHHPH03}
		J. \v{R}eh\`a\v{c}ek, Z. Hradil, O. Haderka, J. Pe\v{r}ina Jr., and M. Hamar,
		Multiple-photon resolving fiber-loop detector,
		Phys. Rev. A \textbf{67}, 061801(R) (2003).
	\bibitem{WDSBY04}
		E. Waks, E. Diamanti, B. C. Sanders, S. D. Bartlett, and Y. Yamamoto,
		Direct Observation of Nonclassical Photon Statistics in Parametric Down-Conversion,
		Phys. Rev. Lett. \textbf{92}, 113602 (2004).
	\bibitem{ZABGGBRP05}
		G. Zambra, A. Andreoni, M. Bondani, M. Gramegna, M. Genovese, G. Brida, A. Rossi, and M. G. A. Paris,
		Experimental Reconstruction of Photon Statistics without Photon Counting,
		Phys. Rev. Lett. \textbf{95}, 063602 (2005).
	\bibitem{JDC07}
		L. A. Jiang, E. A. Dauler, and J. T. Chang,
		Photon-number-resolving detector with 10bits of resolution,
		Phys. Rev. A \textbf{75}, 062325 (2007).
	\bibitem{BGGMPTPOP11}
		G. Brida, M. Genovese, M. Gramegna, A. Meda, F. Piacentini, P. Traina, E. Predazzi, S. Olivares, and M. G. A. Paris,
		Quantum State Reconstruction Using Binary Data from On/Off Photodetection,
		Adv. Sci. Lett. \textbf{4}, 1 (2011).
	\bibitem{MMDL12}
		P.-A. Moreau, J. Mougin-Sisini, F. Devaux, and E. Lantz,
		Realization of the purely spatial Einstein-Podolsky-Rosen paradox in full-field images of spontaneous parametric down-conversion,
		Phys. Rev. A \textbf{86}, 010101(R) (2012).
	\bibitem{FJPF03}
		M. J. Fitch, B. C. Jacobs, T. B. Pittman, and J. D. Franson,
		Photon-number resolution using time-multiplexed single-photon detectors,
		Phys. Rev. A \textbf{68}, 043814 (2003).
	\bibitem{BDFL08}
		J.-L. Blanchet, F. Devaux, L. Furfaro, and E. Lantz,
		Measurement of Sub-Shot-Noise Correlations of Spatial Fluctuations in the Photon-Counting Regime,
		Phys. Rev. Lett. \textbf{101}, 233604 (2008).
	\bibitem{SVA12_85}
		J. Sperling, W. Vogel, and G. S. Agarwal,
		True photocounting statistics of multiple on-off detectors,
		Phys. Rev. A \textbf{85}, 023820 (2012).
	\bibitem{ABA10}
		A. Allevi, M. Bondani, and A. Andreoni,
		Photon-number correlations by photon-number resolving detectors,
		Opt. Lett. \textbf{35}, 1707 (2010).
	\bibitem{ALCS10}
		M. Avenhaus, K. Laiho, M. V. Chekhova, and C. Silberhorn,
		Accessing Higher Order Correlations in Quantum Optical States by Time Multiplexing,
		Phys. Rev. Lett. \textbf{104}, 063602 (2010).
	\bibitem{DYSTS11}
		J. F. Dynes, Z. L. Yuan, A. W. Sharpe, O. Thomas, and A. J. Shields,
		Probing higher order correlations of the photon field with photon number resolving avalanche photodiodes,
		Opt. Express \textbf{19}, 13268 (2011).
	\bibitem{HSRHMSS14}
		G. Harder, C. Silberhorn, J. Rehacek, Z. Hradil, L. Motka, B. Stoklasa, and L. L. S\'anchez-Soto,
		Time-multiplexed measurements of nonclassical light at telecom wavelengths,
		Phys. Rev. A \textbf{90}, 042105 (2014).
	\bibitem{AJBGPHB14}
		A. Allevi, O. Jedrkiewicz, E. Brambilla, A. Gatti, J. Pe\v{r}ina, Jr., O. Haderka, and M. Bondani,
		Coherence properties of high-gain twin beams,
		Phys. Rev. A \textbf{90}, 063812 (2014).
	\bibitem{SBVHBAS15}
		J. Sperling, M. Bohmann, W. Vogel, G. Harder, B. Brecht, V. Ansari, and C. Silberhorn,
		Uncovering Quantum Correlations with Time-Multiplexed Click Detection,
		Phys. Rev. Lett. \textbf{115}, 023601 (2015).
	\bibitem{SVA15}
		J. Sperling, W. Vogel, and G. S. Agarwal,
		Balanced homodyne detection with on-off detector systems: observable nonclassicality criteria,
		Europhys. Lett. \textbf{109}, 34001 (2015).
	\bibitem{LSV15}
		A. Luis, J. Sperling, and W. Vogel,
		Nonlcassicality Phase-Space Functions: More Insight with Fewer Detectors,
		Phys. Rev. Lett. \textbf{114}, 103602 (2015).
	\bibitem{STG08}
		A. A. Semenov, A. V. Turchin, and H. V. Gomonay,
		Detection of quantum light in the presence of noise,
		Phys. Rev. A \textbf{78}, 055803 (2008); \textbf{79}, 019902(E) (2009).
	\bibitem{ASSBW03}
		D. Achilles, C. Silberhorn, C. \'Sliwa, K. Banaszek, and I. A. Walmsley,
		Fiber-assisted detection with photon number resolution,
		Opt. Lett. \textbf{28}, 2387 (2003).
	\bibitem{LFCPSREPW09}
		J. S. Lundeen, A. Feito, H. Coldenstrodt-Ronge, K. L. Pregnell, C. Silberhorn, T. C. Ralph, J. Eisert, M. B. Plenio, and I. A. Walmsley,
		Tomography of quantum detectors,
		Nat. Phys. \textbf{5}, 27 (2009).
	\bibitem{FLCEP09}
		A. Feito, J. S. Lundeen, H. Coldenstrodt-Ronge, J. Eisert, M. B. Plenio, and I. A. Walmsley,
		Measuring measurement: theory and practice,
		New J. Phys. \textbf{11}, 093038 (2009).
	\bibitem{SVA13}
		J. Sperling, W. Vogel, and G. S. Agarwal,
		Correlation measurements with on-off detectors,
		Phys. Rev. A \textbf{88}, 043821 (2013).
	\bibitem{DMM74}
		V. V. Dodonov, I. A. Malkin, and  V. I. Man'ko,
		Even and odd coherent states and excitations of a singular oscillator,
		Physica \textbf{72}, 597 (1974).
	\bibitem{SVA12_109}
		J. Sperling, W. Vogel, and G.S. Agarwal,
		Sub-Binomial Light,
		Phys. Rev. Lett. \textbf{109}, 093601 (2012).
	\bibitem{BDJDBW13}
		T. J. Bartley, G. Donati, X.-M. Jin, A. Datta, M. Barbieri, and I. A. Walmsley,
		Direct Observation of Sub-Binomial Light,
		Phys. Rev Lett. \textbf{110}, 173602 (2013).
	\bibitem{WV96}
		S. Wallentowitz and W. Vogel,
		Unbalanced homodyning for quantum-state measurements,
		Phys. Rev. A \textbf{53}, 4528 (1996).
	\bibitem{HPHP05}
		O. Haderka, J. Pe\v{r}ina, Jr., M. Hamar, and J. Pe\v{r}ina,
		Direct measurement and reconstruction of nonclassical features of twin beams generated in spontaneous parametric down-conversion,
		Phys. Rev. A \textbf{71}, 033815 (2005).
	\bibitem{PHMH12}
		J. Pe\v{r}ina, Jr., M. Hamar, V. Mich\'alek, and O. Haderka,
		Photon-number distributions of twin beams generated in spontaneous parametric down-conversion and measured by an intensified CCD camera,
		Phys. Rev. A \textbf{85}, 023816 (2012).
	\bibitem{KLL03}
		Y. Kang, H. X. Lu, and Y.-H. Lo,
		Dark count probability and quantum efficiency of avalanche photodiodes for single-photon detection,
		Appl. Phys. Lett. \textbf{83}, 2955 (2003).
	\bibitem{BPP98}
		S. M. Barnett, L. S. Phillips, and D. T. Pegg,
		Imperfect photodetection as projection onto mixed states,
		Opt. Commun. \textbf{158}, 45 (1998).
	\bibitem{A87}
		G. S. Agarwal,
		Wigner-function Description of Quantum Noise in Interferometers,
		J. Mod. Opt. \textbf{34}, 909 (1987).
	\bibitem{CommentSRUncertainty}
		The Schr\"odinger-Robertson uncertainty relation for two observables $\Delta\hat A=\hat A-\langle\hat A\rangle\hat 1$ and $\Delta\hat B=\hat B-\langle\hat B\rangle\hat 1$ is
		\begin{align*}
			\langle (\Delta\hat A)^2\rangle\langle (\Delta\hat B)^2\rangle
			\geq& |\langle\Delta\hat A\Delta\hat B\rangle|^2
			\\&=\left|\tfrac{\langle\{\Delta\hat A,\Delta\hat B\}\rangle}{2}\right|^2
			{+}\left|\tfrac{\langle[\Delta\hat A,\Delta\hat B]\rangle}{2i}\right|^2,
		\end{align*}
		with $[\Delta\hat A,\Delta\hat B]=[\hat A,\hat B]$ being the commutator and
		$\langle\{\Delta\hat A,\Delta\hat B\}\rangle=\langle\hat A\hat B\rangle+\langle\hat B\hat A\rangle-2\langle\hat A\rangle\langle\hat B\rangle$ denoting the expectation value of the anti-commutator.
	\bibitem{ZVW96}
		A. Zucchetti, W. Vogel, and D.-G. Welsch,
		Quantum-state homodyne measurement with vacuum ports,
		Phys. Rev. A \textbf{54}, 856 (1996).
	\bibitem{SV06}
		E. Shchukin and W. Vogel,
		Universal Measurement of Quantum Correlations of Radiation,
		Phys. Rev. Lett. \textbf{96}, 200403 (2006).
\end{thebibliography}
\end{document}